\documentclass[12pt]{article}

\usepackage{graphicx}
\usepackage{float}
\usepackage{cite}
\usepackage{amsfonts}
\usepackage{amssymb}

\def\gtwid{\mathrel{\raise.3ex\hbox{$>$\kern-.75em\lower1ex\hbox{$\sim$}}}}
\def\ltwid{\mathrel{\raise.3ex\hbox{$<$\kern-.75em\lower1ex\hbox{$\sim$}}}}
\def\square{\kern1pt\vbox{\hrule height 1.2pt\hbox{\vrule width 1.2pt\hskip 3pt
   \vbox{\vskip 6pt}\hskip 3pt\vrule width 0.6pt}\hrule height 0.6pt}\kern1pt}

\begin{document}

\begin{titlepage}

\begin{flushright}
UFIFT-QG-18-05 , CCTP-2019-1
\end{flushright}

\vskip 2cm

\begin{center}
{\bf Non-Gaussianity from Features}
\end{center}

\vskip 1cm

\begin{center}
S. Basu$^{1\star}$, D. J. Brooker$^{2*}$, N. C. Tsamis$^{3\dagger}$ and 
R. P. Woodard$^{1\ddagger}$
\end{center}

\vskip .5cm

\begin{center}
\it{$^{1}$ Department of Physics, University of Florida,\\
Gainesville, FL 32611, UNITED STATES}
\end{center}

\begin{center}
\it{$^{2}$ U.S. Naval Research Laboratory, Code 7160, \\
4555 Overlook Ave., SW, Washington, DC 20375, UNITED STATES}
\end{center}

\begin{center}
\it{$^{3}$ Institute of Theoretical Physics \& Computational Physics, \\
Department of Physics, University of Crete, \\
GR-710 03 Heraklion, HELLAS}
\end{center}

\vspace{1cm}

\begin{center}
ABSTRACT
\end{center}
The strongest non-Gaussianity in single-scalar potential models of
inflation is associated with features in the power spectrum. We stress 
the importance of accurately modelling the expected signal in order for
the standard estimator to minimize contamination by random noise. We 
present explicit formulae which improve on the approximation introduced
by Adshead, Hu, Dvorkin and Peiris. We also compare with a simple,
analytic model of the first feature, and quantify our results using the
correlators of Hung, Fergusson and Shellard.

\begin{flushleft}
\end{flushleft}

\vskip .5cm

\begin{flushleft}
$^{\star}$ e-mail: shinjinibasu@ufl.edu \\
$^{*}$ e-mail: daniel.brooker@nrl.navy.mil \\
$^{\dagger}$ e-mail: tsamis@physics.uoc.gr \\
$^{\ddagger}$ e-mail: woodard@phys.ufl.edu
\end{flushleft}

\end{titlepage}

\section{Introduction}

The prediction of primordial scalar perturbations \cite{Mukhanov:1981xt} 
in single-scalar inflation described by the Lagrangian,
\begin{equation}
\mathcal{L} = \frac{R \sqrt{-g}}{16 \pi G} - \frac12 \partial_{\mu} \varphi
\partial_{\nu} \varphi g^{\mu\nu} \sqrt{-g} - V(\varphi) \sqrt{-g} \; .
\label{single}
\end{equation}
represents the first (and so far only) observed quantum gravitational phenomena
\cite{Woodard:2009ns,Ashoorioon:2012kh,Krauss:2013pha}. It is frustrating
that we do not know the scalar potential $V(\varphi)$, or even if 
single-scalar inflation is correct. It is also frustrating that so little
guidance for fundamental theory is provided by observation. The approximately 
$10^7$ pixels of data from the primordial spectrum \cite{Aghanim:2018eyx}
seem to be well described by just two numbers,
\begin{equation}
\Delta^2_{\mathcal{R}}(k) \simeq A_s \Bigl( \frac{k}{k_*}\Bigr)^{n_s-1} \!\!\! ,
A_s = (2.105 \pm 0.030) \times 10^{-9} \; , \; n_s = 0.9665 \pm 0.0038 
\; , \label{data}
\end{equation} 
where the pivot is $k_* = 0.05~{\rm Mpc}^{-1}$.

If relation (\ref{data}) is correct then we can reconstruct the 
inflationary geometry in terms of $A_s$, $n_s$ and the still unknown 
tensor-to-scalar ratio $r_{*} < 0.07$ \cite{Array:2015xqh}. Expressing the 
first slow roll parameter $\epsilon(n)$ and the Hubble parameter $H(n)$ 
in terms of the number of e-foldings $\Delta n \equiv n - n_*$ since the 
pivot mode experienced horizon crossing, the lowest order slow roll 
approximation gives,
\begin{equation}
\epsilon(n) \simeq \frac{r_*}{16} e^{(1 - n_s) \Delta n} \quad , \quad
H(n) \simeq H_* \exp\Biggl[ -\frac{r_*}{16 (1 \!-\! n_s)} \Bigl(
e^{(1 - n_s) \Delta n} - 1\Bigr) \Biggr] \; , \label{recongeom}
\end{equation} 
where $8\pi G H^2_* \equiv r_* A_s \pi^2/2$. Using the standard procedure
for reconstructing the inflaton and its potential \cite{Tsamis:1997rk,
Saini:1999ba,Padmanabhan:2002cp,Nojiri:2005pu,Woodard:2006nt,Guo:2006ab} 
we find,
\begin{eqnarray}
\sqrt{8\pi G} \, \Bigl( \varphi(n) - \varphi_*\Bigr) & \equiv & \Delta \psi
\simeq  -\frac1{1 \!-\! n_s} \sqrt{ \frac{r_*}{2}} \Biggr[ e^{\frac12 
(1 - n_s) \Delta n} - 1 \Biggr] \; , \qquad \label{inflation} \\
(8\pi G)^2 V(\varphi) & \simeq & \frac{3}{2} \pi^2 r_* A_s \exp\Biggl[
\sqrt{\frac{r_*}{8}} \, \Delta \psi - \Bigl(\frac{1 \!-\! n_s}{4}\Bigr)
\Delta \psi^2\Biggr] \; . \qquad \label{potential}
\end{eqnarray}

Nature is under no compulsion to comply with human aesthetic prejudices, so 
the featureless, gently sloping potential (\ref{potential}) may be all there 
is to primordial inflation. However, it raises severe issues with the 
fine-tuning of initial conditions needed to make inflation start, and with 
the tendency for small fluctuations to produce dramatically different 
conditions in distant portions of the universe \cite{Ijjas:2013vea}. What 
to make of this has provoked controversy even among some of the pioneers of 
inflation \cite{Guth:2013sya,Linde:2014nna,Ijjas:2014nta}.

The power spectrum data \cite{Ade:2015lrj} actually provide marginal evidence 
for more structure in the form of ``features''. These are transient fluctuations
away from the best fit --- usually a depression of power followed by an excess
at smaller angular scales --- which are visible in the Planck residuals for
$20 \ltwid \ell \ltwid 1500$ \cite{Torrado:2016sls}. These were first noticed in 
WMAP data \cite{Martin:2006rs,Covi:2006ci, Hamann:2007pa} and have persisted 
\cite{Hazra:2014goa,Hazra:2016fkm}. None of the observed features reaches the 
$5\sigma$ level of a detection, but it is conceivable that this threshold might 
be reached by correlating them with other data sets \cite{Achucarro:2012fd}.
We have suggested the possibility of doing this (in the far future) with data 
from the tensor power spectrum \cite{Brooker:2016imi,Brooker:2017kjd}. Here 
we study the prospects for exploiting non-Gaussianity.

Maldacena's analysis \cite{Maldacena:2002vr} established that single-scalar 
inflation (\ref{single}) cannot produce a detectable level of non-Gaussianity if 
the potential is smooth like (\ref{potential}). The effect from a smooth potential 
is widely distributed over the angular bi-spectrum so the standard estimators 
average over all possible 3-point correlators in order to maximize the signal 
\cite{Fergusson:2006pr,Fergusson:2008ra}. Planck has not seen a statistically 
significant indication of non-Gaussianity using any of these standard estimators 
\cite{Ade:2015ava}. On the other hand, it has long been recognized that much 
stronger transient effects can come from features \cite{Adshead:2011bw,
Adshead:2011jq,Hazra:2012yn}. Because these effects are concentrated at certain 
angular scales the standard estimators do not resolve them well. An approximate
computation of the effect from the first feature indicated that its non-Gaussian 
signal is not detectable \cite{Adshead:2011bw}. We will re-examine this problem 
using some recently developed improvements in approximating the scalar mode 
functions \cite{Brooker:2017kjd,Brooker:2017kij}, which unfortunately do not
alter the previous conclusion.

This paper consists of five sections, of which this Introduction is the first. 
Section 2 is devoted to notation and conventions. The various contributions to
non-Gaussianity are listed there, and the one associated with features is
identified. In section 3 we apply our approximation for the scalar mode function 
to derive an analytic expression for the bi-spectrum as a functional of the
inflationary geometry. Section 4 optimizes the parameters for a simple model of 
the first feature in which the bi-spectrum can be computed exactly. Our 
conclusions comprise section 5.

\section{Notation and Conventions}

Our purpose is to elucidate how quantities depend {\it functionally} on the
geometry of inflation. We employ the Hubble representation \cite{Liddle:1994dx} 
using Hubble parameter $H$ and first slow roll parameter $\epsilon$ of the 
homogeneous, isotropic and spatially flat background geometry of inflation,
\begin{equation}
ds^2 = -dt^2 + a^2(t) d\vec{x} \!\cdot\! d\vec{x} \quad \Longrightarrow \quad 
H \equiv \frac{\dot{a}}{a} > 0 \quad , \quad \epsilon \equiv 
-\frac{\dot{H}}{H^2} < 1 \; . \label{geometry}
\end{equation}
It is convenient to regard our time variable as $n \equiv \ln[a(t)/a_i]$, 
the number of e-foldings from the {\it beginning} of inflation. If inflation
ends after $n_e$ e-foldings then the more familiar number of e-foldings {\it 
until} the end of inflation is $N \equiv n_e - n$. With this time variable 
$\epsilon(n)$ provides the simplest representation for the geometry of inflation 
with the Hubble parameter evolved from its initial value $H_i$,
\begin{equation}
H(n) = H_i \exp\Bigl[-\!\int_0^n \!\!\! dm \, \epsilon(m)\Bigr] \; . \label{Hubble}
\end{equation}
We use a prime to denote differentiation with respect to $n$, as in $\epsilon
= -H'/H$.

The key unknown in computing both the scalar power spectrum and the bi-spectrum
is the scalar mode function $v(n,k)$. In our notation its equation, Wronskian
normalization and asymptotically early time form are \cite{Mukhanov:1990me,
Liddle:1993fq},
\begin{equation}
v'' + \Bigl(3 - \epsilon + \frac{\epsilon'}{\epsilon}\Bigr) v' + 
\frac{k^2 v}{H^2 a^2} = 0 \; , \; v v^{\prime *} - v' v^* =
\frac{i}{\epsilon H a^3} \; , \; v \longrightarrow 
\frac{\exp[-i k \! \int_0^{n} \! \frac{dm}{H a}]}{\sqrt{2 k \epsilon a^2}} . 
\label{modeeqn}
\end{equation}
Let $n_k$ stand for the e-folding of first horizon crossing, when modes of wave
number $k$ obey $k \equiv H(n_k) a(n_k)$. One can see from (\ref{modeeqn}) that
the mode function rapidly approaches a constant after this time. The scalar power
spectrum is computed by evolving $v(n,k)$ from its early time form to this
constant,
\begin{equation}
\Delta^2_{\mathcal{R}}(k) = 4\pi G \!\times\! \frac{k^3}{2 \pi^2} \!\times\!
\Bigl\vert v(n,k)\Bigr\vert^2_{n \gg n_k} \; . \label{spectrum}
\end{equation}

Maldacena's expression for the bi-spectrum \cite{Maldacena:2002vr} 
can be expressed as the sum of seven contributions, of which three pairs are 
usually combined \cite{Hazra:2012yn}. In our notation the $I = 1, . . . 7$
contributions each take the form,
\begin{eqnarray}
\lefteqn{B_I(k_1,k_2,k_3) = (4\pi G)^2 \, {\rm Re}\Biggl[ v(n_e,k_1) 
v(n_e,k_2) v(n_e,k_3) } \nonumber \\
& & \hspace{4cm} \times i\!\! \int_0^{n_e} \!\!\! dn \, \epsilon(n) H(n) a^3(n) 
\!\times\! \mathcal{B}^*_{I}(n,k_1,k_2,k_3) \Biggr] \;. \qquad \label{Bform}
\end{eqnarray}
The four unconjugated $\mathcal{B}_I(n,k_1,k_2,k_3)$ combinations are,
\begin{eqnarray}
\mathcal{B}_{1+3} & = & \epsilon \Biggl[ \frac{K^4_{123}}{k^2_2 k^2_3} v_1
v_2' v_3' + \frac{K^4_{231}}{k^2_3 k^2_1} v_1' v_2 v_3' + \frac{K^4_{312}}{
k^2_1 k^2_2} v_1' v_2' v_3 \Biggr] \; , \label {B13} \\
\mathcal{B}_{2} & = & \epsilon \Biggl[ \Bigl( 
\frac{k^2_1 \!+\! k^2_2 \!+\! k^2_3}{H^2 a^2} \Bigr) v_1 v_2 v_3 \Biggr] \; , 
\label{B2} \\
\mathcal{B}_{5+6} & = & \epsilon^2 \Biggl[ \frac{K^4}{k^2_2 k^2_3} v_1 v_2' v_3' + 
\frac{K^4}{k^2_3 k^2_1} v_1' v_2 v_3' + \frac{K^4}{k^2_1 k^2_2} v_1' v_2' v_3 \Biggr] 
\; , \label {B56} \\ 
\mathcal{B}_{4+7} & = & \frac{\epsilon'}{\epsilon} \Biggl[\Bigl( \frac{k^2_1 \!+\! 
k^2_2 \!+\! k^2_3}{H^2 a^2}\Bigr) v_1 v_2 v_3 - v_1 v_2' v_3' - v_1' v_2 v_3'
- v_1' v_2' v_3 \Biggr] \; , \qquad \label{B47}
\end{eqnarray}
where the 4th order momentum factors in (\ref{B13}) and (\ref{B56}) are,
\begin{eqnarray}
K^4_{123} & \equiv & k^2_1 (k^2_2 \!+\! k^2_3) + 2 k^2_2 k^2_3 - (k^2_2 \!-\! 
k^2_3)^2 \; , \label{K123} \\
K^4 & \equiv & k^4_1 + k^4_2 + k^4_3 - 2 k^2_1 k^2_2 - 2 k^2_2 k^2_3 - 
2 k^2_3 k^2_1 \; . \label{K4}
\end{eqnarray}

Two things are apparent from the initial factors of $\epsilon$ in 
expressions (\ref{B13}-\ref{B47}). First, non-Gaussianity is small for 
smooth potentials like (\ref{potential}) because $\epsilon$ is small and 
varies slowly. From (\ref{recongeom}) we see that $\epsilon \sim \frac1{16} 
r_* < 0.0044$, and even the factor of $\epsilon'/\epsilon$ in (\ref{B47}) 
is $1 - n_s \sim 0.034$. Second, much larger non-Gaussianity can arise from
$\mathcal{B}_{4+7}$ in models with features. In that case $\epsilon$ remains
small, but $\epsilon'/\epsilon$ can reach order one over a small range of $n$.

The mode-dependent factors inside the square brackets of (\ref{B13}-\ref{B47})
are also informative when combined with three insights from the mode equation 
(\ref{modeeqn}):
\begin{enumerate}
\item{The mode function $v(n,k)$ is oscillatory and falling off like $1/a$ 
until it freezes in to a constant $V(k)$ (which might be complex) around $n 
\approx n_k$;}
\item{The approach to $V(k)$ has real part ${\rm Re}[v(n,k)/V(k)] \sim
(k/Ha)^2$; and}
\item{The approach has ${\rm Im}[v(n,k)/V(k)] \sim 
-1/2\epsilon H a^3 \vert V(k)\vert^2$.}
\end{enumerate}
Together with the general form (\ref{Bform}), these facts imply that the 
$n$-integrand for each of the four contributions is oscillatory before the
largest of the three wave numbers has experienced horizon crossing and falls 
off like $1/a^2$ thereafter. This has important consequences for designing 
estimators to detect non-Gaussianity. When the potential is smooth both 
$\epsilon(n)$ and $\partial_n \ln[\epsilon]$ are nearly constant, so all 
wave numbers will show nearly the same effect and the best strategy is to 
combine them as the standard estimators do. However, when a feature is
present the factor of $\partial_n \ln[\epsilon(n)]$ in (\ref{B47}) becomes
significant in small range of $n$, and the non-Gaussian signal will be much
larger for modes which experience horizon crossing around that time. 
Averaging over all observable wave numbers runs the risk of drowning a
real signal in noise.

Because conventions differ we close by reviewing how the fundamental fields 
relate to $\Delta^2_{\mathcal{R}}(k)$ and $B(k_1,k_2,k_3)$. We use the gauge 
of Salopek, Bond and Bardeen \cite{Salopek:1988qh} in which time is fixed by
setting the inflaton to its background value and the graviton field is 
transverse. In this gauge the metric components $g_{00}$ and $g_{0i}$ are
constrained and the dynamical variables $\zeta(n,\vec{x})$ and $h_{ij}(n,\vec{x})$ 
reside in the spatial components,
\begin{equation}
g_{ij}(n,\vec{x}) = a^2 e^{2 \zeta(n,\vec{x})} \!\times\! \Bigl[ e^{h(n,\vec{x})}
\Bigr]_{ij} \qquad , \qquad h_{ii}(n,\vec{x}) = 0 \; . \label{zetahij}
\end{equation}
Scalar perturbations derive from $\zeta(n,\vec{x})$ whose free field expansion 
is,
\begin{equation}
\widetilde{\zeta}(n,\vec{k}) \equiv \int \!\! d^3x \, e^{-i \vec{k} \cdot \vec{x}} 
\zeta(n,\vec{x}) = \sqrt{4 \pi G} \, \Bigl[ v(n,k) \alpha(\vec{k}) + v^*(n,k) 
\alpha^{\dagger}(-\vec{k}) \Bigr] \; , \label{freezeta}
\end{equation}
where $\alpha^{\dagger}$ and $\alpha$ are creation and annihilation operators,
\begin{equation}
\Bigl[ \alpha(\vec{k}) , \alpha^{\dagger}(\vec{p}) \Bigr] = (2\pi)^3 
\delta^3(\vec{k} \!-\! \vec{p}) \qquad , \qquad \alpha(\vec{k}) \Bigl\vert
\Omega \Bigr\rangle = 0 \; . \label{opsvac}
\end{equation}
Assuming the wave numbers experience horizon crossing before the end of inflation
$n_e$, our power spectrum and bi-spectrum are,
\begin{eqnarray}
\Bigl\langle \Omega \Bigl\vert \widetilde{\zeta}(n_e,\vec{k}) 
\widetilde{\zeta}(n_e,\vec{p}) \Bigr\vert \Omega \Bigr\rangle & \!\!\!=
\!\!\! & \frac{2 \pi^2}{k^3} \!\times\! \Delta^2_{\mathcal{R}}(k) \!\times\!
(2\pi)^3 \delta^3(\vec{k} \!+\! \vec{p} \,) \; , \label{zeta2} \\
\Bigl\langle \Omega \Bigl\vert \widetilde{\zeta}(n_e,\vec{k}_1) 
\widetilde{\zeta}(n_e,\vec{k}_2) \widetilde{\zeta}(n_e,\vec{k}_3) \Bigr\vert 
\Omega \Bigr\rangle & \!\!\!=\!\!\! & B(k_1,k_2,k_3) \!\times\! (2\pi)^3 
\delta^3(\vec{k}_1 \!+\! \vec{k}_2 \!+\! \vec{k}_3) \; . \qquad \label{zeta3}
\end{eqnarray}
Note that while the power spectrum is dimensionless, the bi-spectrum has the
dimension of $k^6$.

\section{Analytic Approximation for the Bi-Spectrum}

In this section we first convert the key contribution (\ref{B47}) from the
mode function $v(n,k)$ to its norm-square $N(n,k)$. Then we introduce an
approximation \cite{Brooker:2017kjd,Brooker:2017kij} which should be very 
accurate for the physically relevant case of small $\epsilon(n)$ but 
significant $\partial_n \ln[\epsilon(n)]$. Finally, we study a model of 
the first feature to compare our result for $B_{4+7}(k_1,k_2,k_3)$ with the
simpler approximation of Adshead, Dvorkin, Hu and Peiris \cite{Adshead:2011bw}. 

\subsection{Approximating the mode functions}

Even considered as a purely numerical problem, it is better to convert the
equations (\ref{modeeqn}) for $v(n,k)$ into relations for $N(n,k) \equiv 
\vert v(n,k)\vert^2$ \cite{Romania:2012tb}. Avoiding the need to keep
track of the phase makes about a quadratic improvement in convergence.
Further, nothing is lost because the phase can be recovered by a simple 
integration \cite{Brooker:2017kjd},
\begin{equation}
v(n,k) = \sqrt{N(n,k)} \, \exp\Bigl[-i \! \int_0^{n} \frac{dm}{2 
\epsilon H a^3 N} \Bigr] \equiv \sqrt{N(n,k)} \, e^{i \theta(n,k)} \; . 
\label{Ntov}
\end{equation}

It is best to begin with the outer factors of $v(n_e,k)$ in expression
(\ref{Bform}). Assuming the various wave numbers have experienced 
horizon crossing these outer mode functions can be expressed in terms of 
the power spectrum (\ref{spectrum}),
\begin{equation}
v(n_e,k) = \sqrt{\frac{\pi}{2 G k^3}} \, \Delta_{\mathcal{R}}(k) 
e^{i \theta(n_e,k)} \; . \label{outermodes}
\end{equation}
We next combine each outer phase with the appropriate inner phase,
\begin{equation}
\widehat{v}(n,k) \equiv v(n,k) e^{-i \theta(n_e,k)} \Longrightarrow 
\theta(n,k) - \theta(n_e,k) = \!\int_{n}^{n_e} \!\! \frac{dm}{\epsilon H a^3
N} \equiv \phi(n,k) \; . \label{phidef}
\end{equation}
Note that $\phi(n,k)$ approaches zero like $1/a^3$ for large $n$. At
this stage one can recognize the real part of the undifferentiated terms,
\begin{equation}
{\rm Re}\Bigl[i \, \widehat{v}_1^* \, \widehat{v}_2^* \, \widehat{v}_3^*\Bigr] 
= \sqrt{N_1 N_2 N_3} \, \sin(\phi_1 \!+\! \phi_2 \!+\! \phi_3) \; .
\label{undiffterms}
\end{equation}
The differentiated terms are more complicated,
\begin{equation}
\widehat{v}'(n,k) = \widehat{v}(n,k) \Bigl[ \frac{N'(n,k)}{2 N(n,k)} + 
i \phi'(n,k) \Bigr] \; . \label{vprime}
\end{equation}
Hence we have,
\begin{eqnarray}
\lefteqn{{\rm Re}\Bigl[i \, \widehat{v}_1^* \, \widehat{v}_2^{\prime *} \, 
\widehat{v}_3^{\prime *} \Bigr] = \sqrt{N_1 N_2 N_3} \Biggl\{
\sin(\phi_1 \!+\! \phi_2 \!+\! \phi_3) \Bigl[ \frac{N_2'}{2 N_2} \frac{N_3'}{2 N_3}
- \phi_2' \phi_3'\Bigr] } \nonumber \\
& & \hspace{5cm} + \cos(\phi_1 \!+\! \phi_2 \!+\! \phi_3) \Bigl[
\frac{N_2'}{2 N_2} \phi_3' + \frac{N_3'}{2 N_3} \phi_2'\Bigr] \Biggr\} . \qquad
\label{deriterm}
\end{eqnarray}
There are three terms such as (\ref{deriterm}), so putting everything together
gives,
\begin{eqnarray}
\lefteqn{B_{4+7}(k_1,k_2,k_3) = \frac{4 \pi^4 \Delta_{\mathcal{R}}(k_1) 
\Delta_{\mathcal{R}}(k_2) \Delta_{\mathcal{R}}(k_3)}{k_1^2 \, k_2^2 \, k_3^2} \!\times\!
\sqrt{ \frac{2 G k_1 k_2 k_3}{\pi}} \int_0^{n_e} \!\!\!\! dn \, \epsilon' H a^3 }
\nonumber \\
& & \hspace{-.5cm} \times \sqrt{N_1 N_2 N_3} \Biggr\{\sin(\phi_1 \!+\! \phi_2 \!+\! 
\phi_3) \Biggl[ \Bigl( \frac{k_1^2 \!+\! k_2^2 \!+\! k_3^2}{H^2 a^2} \Bigr) \!-\! 
\frac{N_2'}{2 N_2} \frac{N_3'}{2 N_3} \!+\! \phi_2' \phi_3' - \dots\Biggr] \nonumber \\
& & \hspace{0cm} - \cos(\phi_1 \!+\! \phi_2 \!+\! \phi_3) \Bigl[ \frac{N_1'}{2 N_1}
(\phi_2' \!+\! \phi_3') +  \frac{N_2'}{2 N_2} (\phi_3' \!+\! \phi_1') +  
\frac{N_3'}{2 N_3} (\phi_1' \!+\! \phi_2') \Bigr] \Biggr\} . \qquad \label{fullB47}
\end{eqnarray}

To develop a useful approximation for (\ref{fullB47}) we first factor $N(n,k)$ into 
the instantaneously constant $\epsilon$ solution $N_0(n,k)$ times the exponential of 
a residual $g(n,k)$ which is sourced by derivatives of $\ln[\epsilon(n)]$ 
\cite{Brooker:2017kjd,Brooker:2017kij},
\begin{equation}
N(n,k) = N_0(n,k) \times \exp\Bigl[ -\frac12 g(n,k)\Bigr] \; . \label{gdef}
\end{equation}
Of course the derivatives of $\ln[\epsilon(n)]$ which source $g(n,k)$ are of great 
concern in the study of features, as is the potentially large factor of $1/\epsilon$ 
in $N_0(n,k)$. Taking all the other factors of $\epsilon$ to zero causes a
negligible loss of accuracy. The resulting approximation involves three functions 
$\widetilde{g}(n,n_k)$, $\widetilde{\gamma}'(n,n_k)$ and $\widetilde{\phi}(n,n_k)$ 
which must be tabulated over a narrow range of $n$ and $n_k$,
\begin{eqnarray}
\lefteqn{\widetilde{B}_{4+7}(k_1,k_2,k_3) = \frac{4 \pi^4 \Delta_{\mathcal{R}}(k_1) 
\Delta_{\mathcal{R}}(k_2) \Delta_{\mathcal{R}}(k_3)}{k_1^2 \, k_2^2 \, k_3^2} \!\times\!
-\! \int_0^{n_e} \!\!\!\! dn \, \partial_n \sqrt{\frac{G H^2(n)}{\pi \epsilon(n)}} }
\nonumber \\
& & \hspace{-.5cm} \times \sqrt{(1 \!+\! e^{2 \Delta n_1}) (1 \!+\! e^{2 \Delta n_2})
(1 \!+\! e^{2 \Delta n_3})} \, e^{-\frac12 ( \widetilde{g}_1 + \widetilde{g}_2 +
\widetilde{g}_3)} \Biggl\{ \sin(\widetilde{\phi}_1 \!+\! \widetilde{\phi}_2 \!+\!
\widetilde{\phi}_3) \nonumber \\
& & \hspace{0cm} \times \Biggl[ e^{-2\Delta n_1} \!-\! \Bigl(\frac1{1 \!+\! 
e^{2 \Delta n_2}} \!+\! \frac14 \widetilde{\gamma}'_2 \Bigr) \Bigl(\frac1{1 \!+\! 
e^{2 \Delta n_3}} \!+\! \frac14 \widetilde{\gamma}'_3\Bigr) \!+\! \widetilde{\phi}'_2 
\widetilde{\phi}'_3 + (231) + (312) \Biggr] \nonumber \\
& & \hspace{.5cm} - \cos(\widetilde{\phi}_1 \!+\! \widetilde{\phi}_2 \!+\! 
\widetilde{\phi}_3) \Biggl[ \Bigl(\frac1{1 \!+\! e^{2 \Delta n_1}} \!+\! \frac14 
\widetilde{\gamma}'_1 \Bigr) (\widetilde{\phi}_2' \!+\! \widetilde{\phi}_3') + (231)
+ (312) \Biggr] \Biggr\} . \label{approxB47} \qquad 
\end{eqnarray}
Here and henceforth $\Delta n_i \equiv n - n_i$, where $n_i$ is the e-folding at 
which wave number $k_i$ experiences horizon crossing.

The tabulated function $\widetilde{g}(n,n_k)$ represents an approximation of the
amplitude residual $g(n,k)$ in (\ref{gdef}). It is expressed as a Green's function 
integral over sources before and after horizon crossing,
\begin{eqnarray}
S_b(m) & = & \partial^2_m \ln[ \epsilon(m)] + \frac12 \Bigl( \partial_m \ln[\epsilon(m)]
\Bigr)^2 + 3 \partial_m \ln[\epsilon(m)] \; , \label{Sbefore} \\
S_a(m,n_k) & = & \frac{2 \partial_m \ln[\epsilon(m)]}{1 \!+\! e^{2 \Delta m}} + 
\Biggl( \frac{2 e^{-\Delta m} \frac{\epsilon(n_k)}{\epsilon(m)} }{1 \!+\! e^{2 \Delta m}}
\Biggr)^2 \; , \label{Safter}
\end{eqnarray}
where $\Delta m \equiv m - n_k$. The integral expression for $\widetilde{g}(n,n_k)$ is,
\begin{eqnarray}
\lefteqn{\widetilde{g}(n,n_k) = -2\theta(-\Delta n) \! \int_0^{n} \!\!\! dm \, 
G(\Delta m,\Delta n) S_b(m) +2 \theta(\Delta n) \Biggl\{ G(0,\Delta n) 
\frac{\epsilon'(n_k)}{\epsilon(n_k)} } \nonumber \\
& & \hspace{1cm} - \int_0^{n_k} \!\!\! dm \, G(\Delta m,\Delta n) S_b(m) 
- \int_{n_k}^{n} \!\!\! dm \, G(\Delta m,\Delta n) S_a(m,n_k)\Biggr\} , \qquad 
\label{gapprox}
\end{eqnarray}
where the Green's function is,
\begin{eqnarray}
\lefteqn{G(\Delta m,\Delta n) = \frac12 \Bigl(e^{\Delta m} \!+\! e^{3 \Delta m}\Bigr)
} \nonumber \\
& & \hspace{1cm} \times \sin\Biggl[2 e^{-\Delta m} - 2 \tan^{-1}\Bigl(e^{-\Delta m}\Bigr) 
- 2 e^{-\Delta n} + 2 \tan^{-1}\Bigl(e^{-\Delta n}\Bigr) \Biggr] . \label{Green} \qquad
\end{eqnarray}
Differentiating the Green's function with respect to $n$ gives,
\begin{eqnarray}
\lefteqn{\partial_n G(\Delta m,\Delta n) = \Biggl(\frac{e^{\Delta m} \!+\! e^{3 \Delta m}}{
e^{\Delta n} \!+\! e^{3 \Delta n} } \Biggr) } \nonumber \\
& & \hspace{1cm} \times \cos\Biggl[2 e^{-\Delta m} - 2 \tan^{-1}\Bigl(e^{-\Delta m}\Bigr) 
- 2 e^{-\Delta n} + 2 \tan^{-1}\Bigl(e^{-\Delta n}\Bigr) \Biggr] . 
\label{Greenprime} \qquad
\end{eqnarray}
It occurs in the second of the tabulated functions,
\begin{equation}
\widetilde{\gamma}'(n,n_k) = 2\theta(-\Delta n) \partial_n \ln[\epsilon(n)] + \partial_n
\widetilde{g}(n,n_k) \; . \label{gprime}
\end{equation}
The final tabulated function is our approximation of the angle $\phi(n,k)$,
\begin{equation}
\widetilde{\phi}(n,n_k) = \int_{n}^{n_e} \!\!\! dm \, \frac{e^{-\Delta m + \frac12 
\widetilde{g}(m,n_k)}}{1 \!+\! e^{2 \Delta m}} \; . \label{phiapprox}
\end{equation}
Note that its derivative does not require separate tabulation,
\begin{equation}
\widetilde{\phi}'(n,n_k) = -\frac{e^{-\Delta n + \frac12 
\widetilde{g}(n,n_k)}}{1 \!+\! e^{2 \Delta n}} \; . \label{phiprime}
\end{equation}

Adshead, Dvorkin, Hu and Peiris \cite{Adshead:2011bw} introduced a much
simpler approximation which, in our language, corresponds to setting
$\widetilde{g}(n,n_k)$ and $\widetilde{\gamma}'(n,k)$ to zero in 
expression (\ref{approxB47}). Note that this reduces the angle and its 
derivative to be functions of just the single variable $\Delta n = n - n_k$,
\begin{equation}
\widetilde{\phi}(n,n_k) \Bigl\vert_{\widetilde{g} = 0} = e^{-\Delta n} -
\tan^{-1}\Bigl(e^{-\Delta n}\Bigr) \quad , \quad 
\widetilde{\phi}'(n,n_k) \Bigl\vert_{\widetilde{g} = 0} = -
\frac{e^{-\Delta n}}{1 \!+\! e^{2 \Delta n}} \; . \label{phiADHP}
\end{equation}
This approximation is certainly simpler to implement, but it completely
ignores how the inner mode functions change in response to the feature. 

\subsection{The Step Model}

The model we shall study belongs to a class introduced in 2001 by Adams, 
Cresswell and Easther \cite{Adams:2001vc},
\begin{equation}
V(\varphi) = \frac12 m^2 \varphi^2 \times \Bigl[1 + c \, 
{\rm tanh}\Bigl( \frac{\varphi \!-\! b}{d}\Bigr) \Bigr] \; . \label{Step1}
\end{equation}
A fit to the first feature ($20 \ltwid \ell \ltwid 40)$ using WMAP data
gave \cite{Mortonson:2009qv},
\begin{equation}
b = \frac{14.668}{\sqrt{8\pi G}} \;\; , \; c = 1.505 \!\times\! 10^{-3} 
\;\; , \; d = \frac{0.02705}{\sqrt{8\pi G}} \;\; , \; m = \frac{7.126 
\!\times\! 10^{-6}}{\sqrt{8\pi G}} \; . \label{Step2}
\end{equation}
\begin{figure}[H]
\includegraphics[width=6.0cm,height=4.0cm]{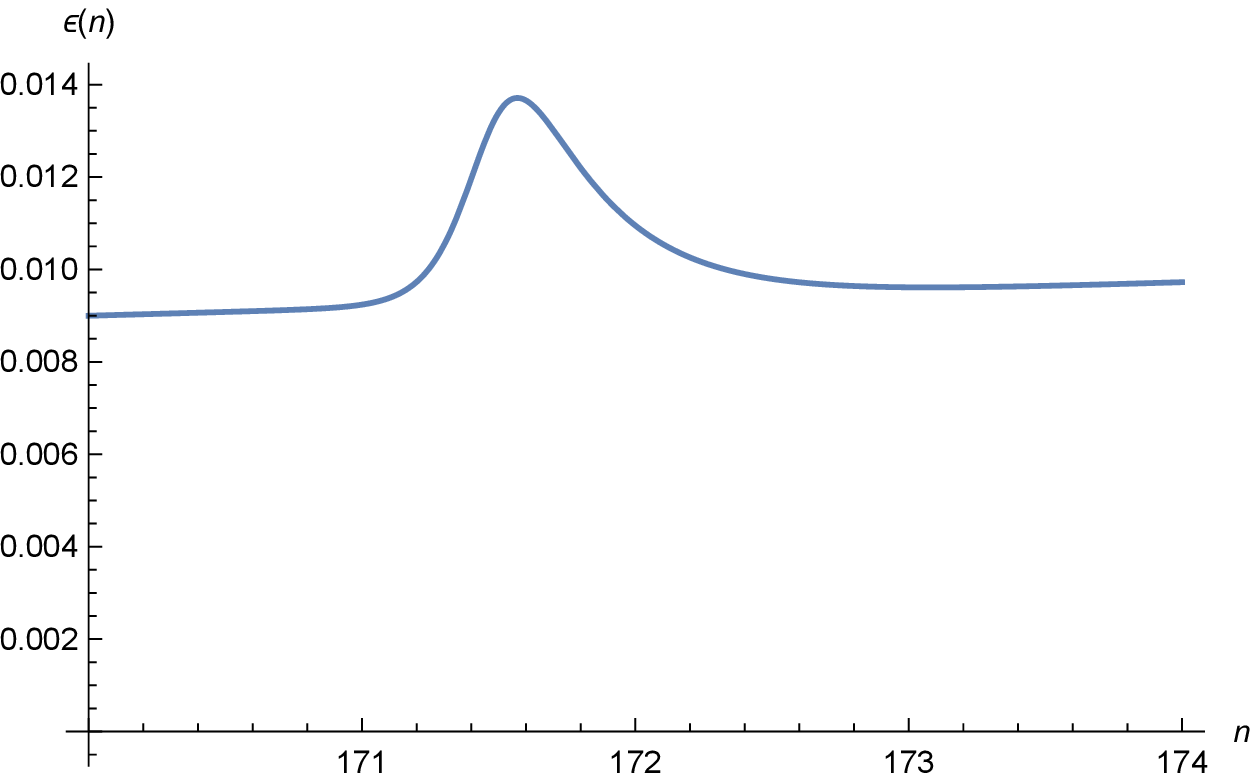}
\hspace{1cm}
\includegraphics[width=6.0cm,height=4.0cm]{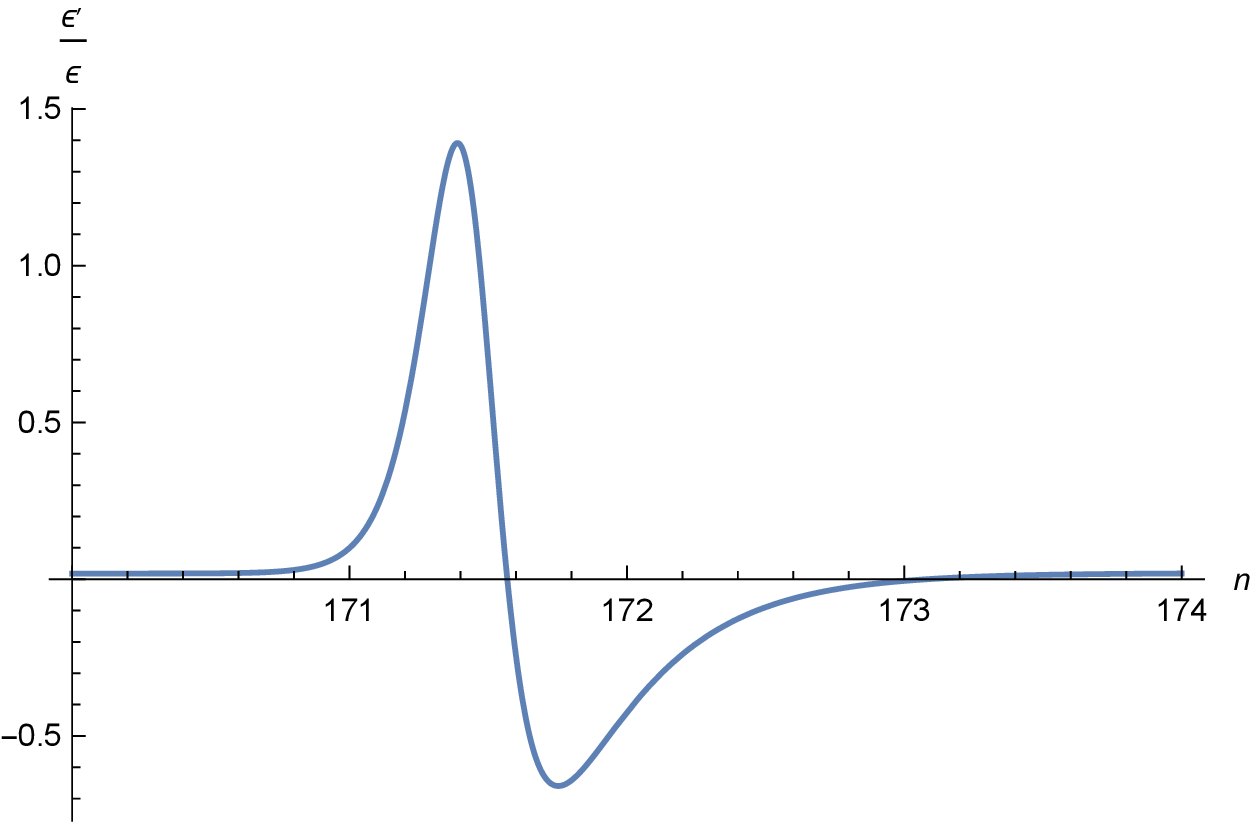}
\caption{\footnotesize The left hand graph gives $\epsilon(n)$ for the Step 
Model (\ref{Step1}-\ref{Step2}). The right hand graph shows $\partial_n \ln[
\epsilon(n)]$ for this model. Note that the logarithmic derivative is 
only significant in the narrow range $170.8 \ltwid n \ltwid 172.8$.}
\label{StepGeometry}
\end{figure}
Figure~\ref{StepGeometry} shows the first slow roll parameter and its
logarithmic derivative for this model. Two obvious points are:
\begin{enumerate}
\item{The first slow roll parameter is always very small;\footnote{
It is actually a little too large for the improved bounds on the 
tensor-to-scalar ratio \cite{Array:2015xqh} since the time of WMAP. However,
the model serves well enough for the purposes of illustration.} and}
\item{The crucial factor of $\partial_{n} \ln[\epsilon(n)]$ which sources
non-Gaussianity is only significant for the two e-foldings $170.8 \ltwid n
\ltwid 172.8$.}
\end{enumerate}
Inflation ends for this model at $n_e \simeq 225.6$ so the feature peaks 
about $54$ e-foldings before the end of inflation.

\begin{figure}[H]
\includegraphics[width=6.0cm,height=4.0cm]{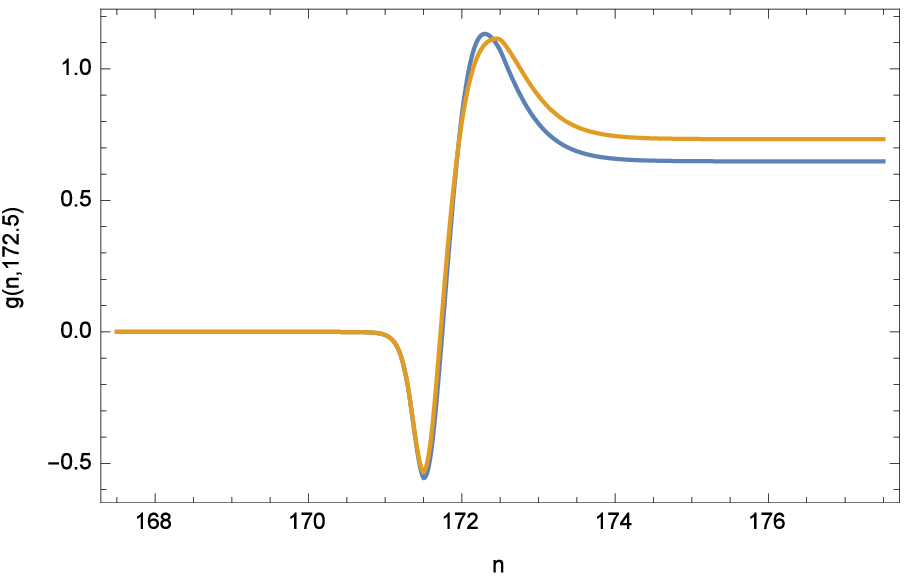}
\hspace{1cm}
\includegraphics[width=6.0cm,height=4.0cm]{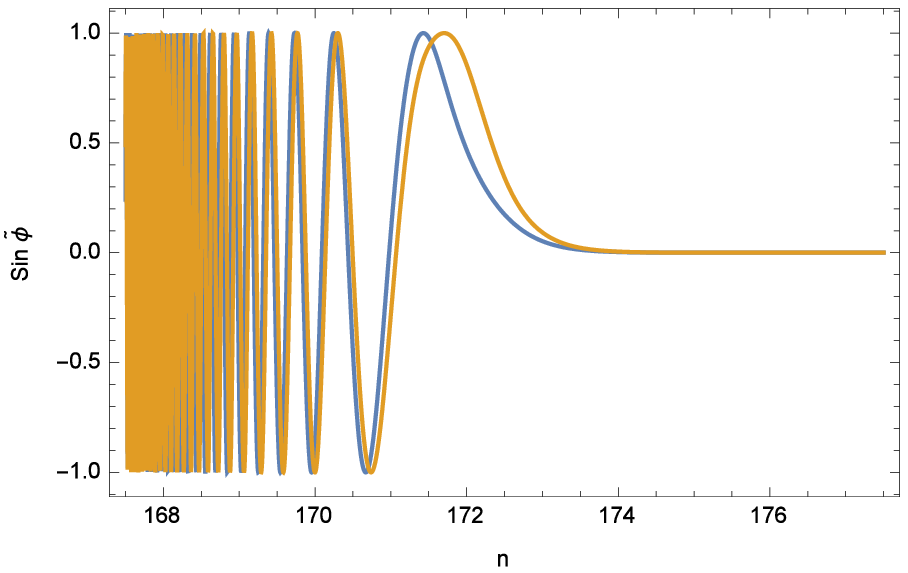}
\caption{\footnotesize Comparison between exact results (in blue) and our 
approximations (in yellow) for the amplitude correction (\ref{gapprox}) and the 
phase (\ref{phiapprox}). The left hand graph shows $g(n,k)$ and the right hand 
graph shows $\sin[\widetilde{\phi}(n,k)]$.} 
\label{n1725}
\end{figure}

Let us first establish that our approximations for the amplitude correction
(\ref{gapprox}) and for the phase (\ref{phiapprox}) are valid. 
Figure~\ref{n1725} displays the exact results (in blue) versus our 
approximations (in yellow) for the case of $n_k = 172.5$ where the amplitude 
correction is close to it maximum. The agreement is good, except for an
offset at late times which is due to $g(n,k)$ having become large enough 
around $n \approx 172$ that nonlinear corrections matter 
\cite{Brooker:2017kij}. For most values of $n_k$ this is not an issue and,
even for $n_k = 172.5$, the rightmost graph of Fig.~\ref{StepGeometry} shows
that the offset has little effect on non-Gaussianity.

\begin{figure}[H]
\includegraphics[width=4.0cm,height=3.0cm]{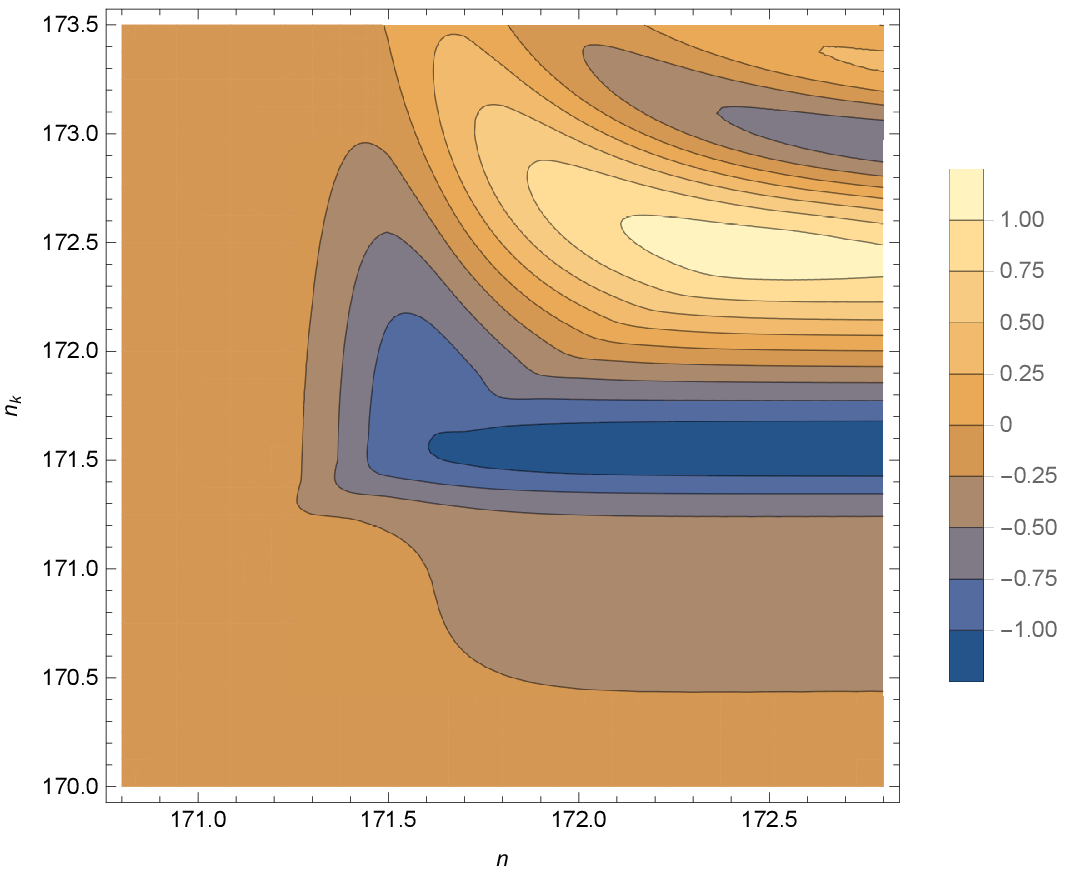}
\hspace{0.5cm}
\includegraphics[width=4.0cm,height=3.0cm]{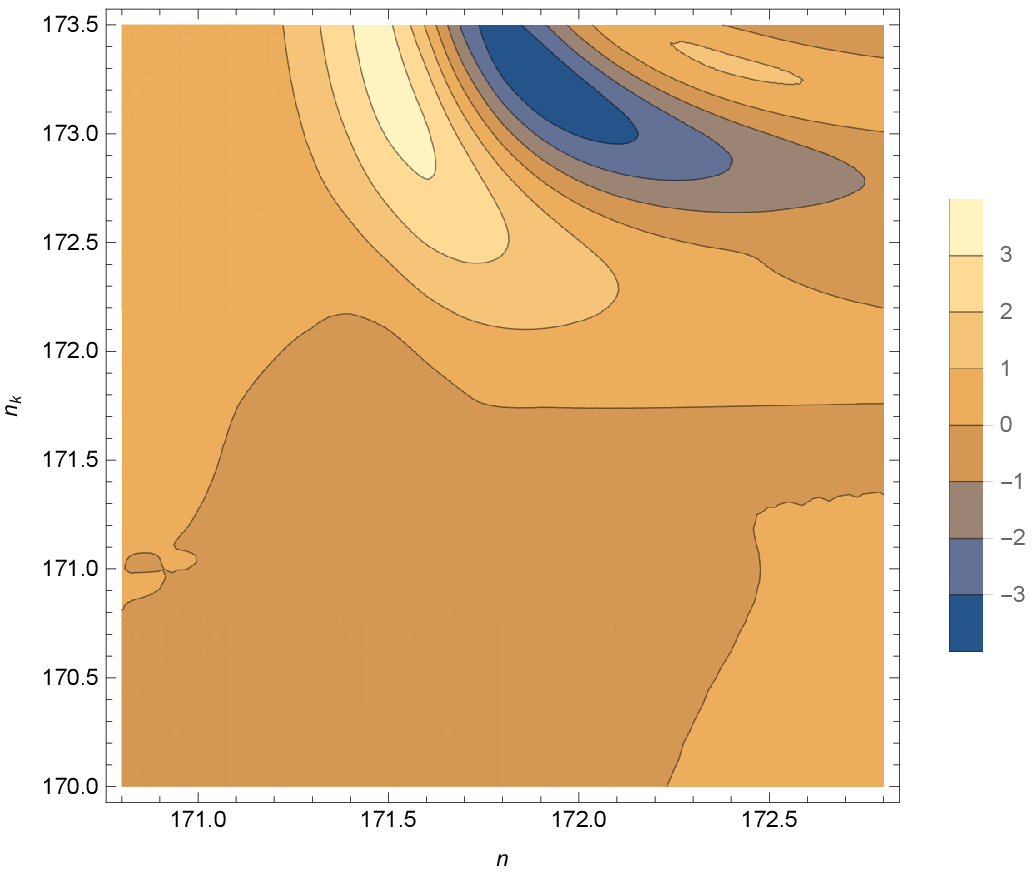}
\hspace{0.5cm}
\includegraphics[width=4.0cm,height=3.0cm]{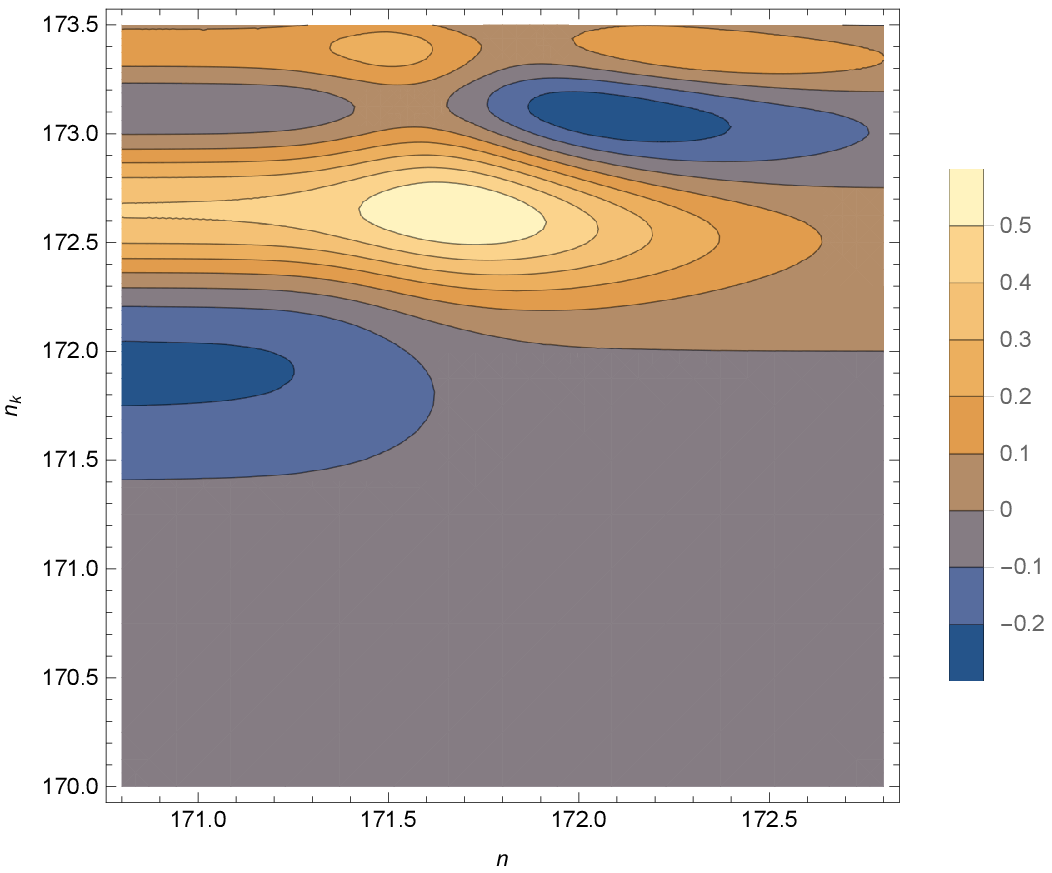}
\caption{\footnotesize Various correction factors for the Step Model. The left 
hand graph gives our approximation (\ref{gapprox}) for ($-4$ times the logarithm 
of) the amplitude correction $g(n,n_k)$ in the Step Model. The middle graph 
shows the derivative factor (\ref{gprime}). And the right hand graph shows
how much our approximation (\ref{phiapprox}) differs from the de Sitter
result (\ref{phiADHP}).}
\label{ModelDeviations}
\end{figure}

In view of point (2) above, we only require the tabulated functions 
$\widetilde{g}(n,n_k)$, $\widetilde{\gamma}'(n,n_k)$ and 
$\widetilde{\phi}(n,nk)$ for the two e-foldings from $n = 170.8$ to
$n = 172.8$. Figure~\ref{ModelDeviations} shows contour plots of these 
functions for modes which experience horizon crossing in the range $170 < 
n_k < 173.5$.
It is important to bear in mind that the source $\partial_n \ln[\epsilon(n)]$
on Fig.~\ref{StepGeometry} modulates how the corrections of 
Fig.~\ref{ModelDeviations} affect non-Gaussianity. So although the graph
of $\widetilde{g}(n,n_k)$ shows a strong amplitude enhancement for $n_k
\simeq 171.5$, and an equally strong suppression for $n_k \simeq 172.5$,
the latter effect is much less significant because it peaks for $n \gtwid
172.1$, by which point $\partial_n \ln[\epsilon(n)]$ is small. Because of 
this modulation, the biggest correction comes from the large positive phase 
shift at $n_k \simeq 172.6$, which peaks at $n \simeq 171.7$. 

\begin{figure}[H]
\includegraphics[width=6.0cm,height=4.0cm]{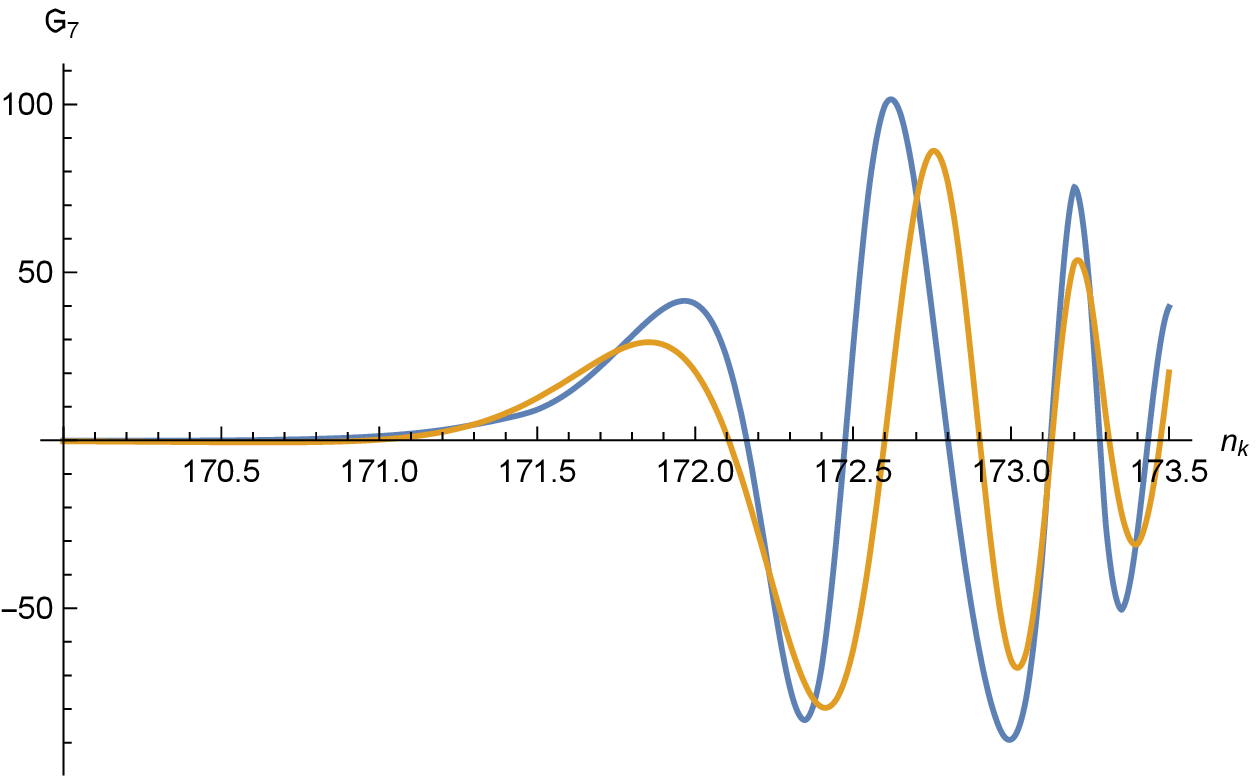}
\hspace{1cm}
\includegraphics[width=6.0cm,height=4.0cm]{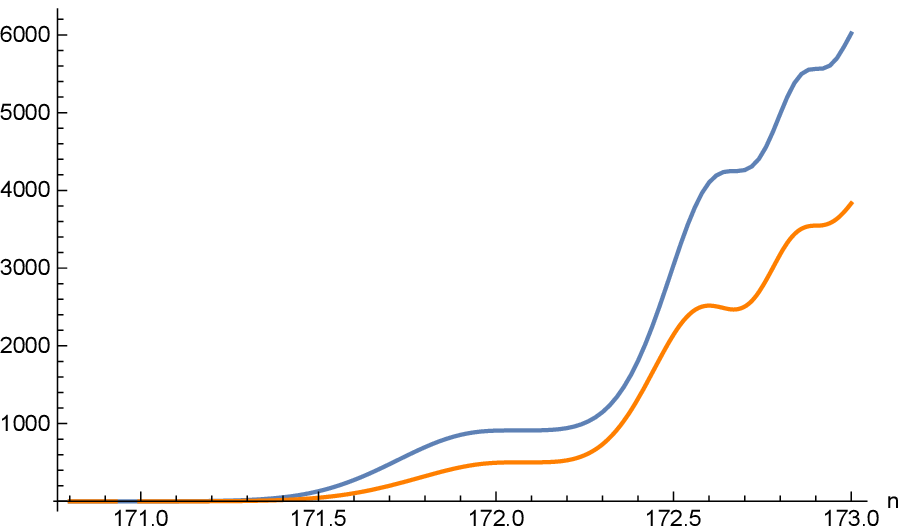}
\caption{\footnotesize The left hand graph shows the ``inner'' part of expression
(\ref{approxB47}), starting from $-\int_0^{n_e} dn \dots$, for the
equilateral triangle case of $k_1 = k_2 = k_3$. The blue curve shows
our approximation while the yellow curve shows the simpler approximation 
of Adshead, Dvorkin, Hu and Peiris \cite{Adshead:2011bw}. The right hand 
graph shows the integral of the square of our approximation (in blue)
versus the product of our approximation times theirs (in yellow). The
ratio of the areas under the yellow to the blue curves is about $0.637$
at the end.}
\label{Compare}
\end{figure}

Figure~\ref{Compare} gives some idea of the significance of the various 
corrections we have introduced to the approximation of Adshead, Dvorkin, 
Hu and Peiris \cite{Adshead:2011bw}, but it is limited by the assumption 
that $k_1 = k_2 = k_3$. The correlators Hung, Fergusson and Shellard 
\cite{Hung:2019ygc} provide a more detailed comparison between any two
bi-spectra $B_i(k_1,k_2,k_3)$ and $B_j(k_1,k_2,k_3)$ which possess the 
same power spectrum $\Delta^2_{\mathcal{R}}(k)$. They are formed from 
ratios of ``inner products'' defined as,
\begin{equation}
[ B_i , B_j] \equiv {\rm Constant} \times \int_{\mathcal{V}_B} \!\!\!\!\!
dk_1 dk_2 dk_3 (k_1 k_2 k_3)^4 \frac{B_i(k_1,k_2,k_3) B_j(k_1,k_2,k_3)}{
\Delta^2_{\mathcal{R}}(k_1) \Delta^2_{\mathcal{R}}(k_2) 
\Delta^3_{\mathcal{R}}(k_3)} \; , \label{inner}
\end{equation}
where $\mathcal{V}_B$ indicates the range of the wave numbers that obey
the triangle condition ($\vert k_1 - k_2\vert < k_3 < k_1 + k_2$), plus
whatever other restrictions we wish to impose, and the multiplicative 
constant is irrelevant. Hung, Fergusson and Shellard use these inner
products to form, respectively, shape, amplitude and total correlators,
\begin{eqnarray}
\mathcal{S}(B_i,B_j) \equiv \frac{ [B_i , B_j]}{\sqrt{ [B_i , B_i]
[B_j , B_j]}} \qquad & , & \qquad
\mathcal{A}(B_i,B_j) \equiv \sqrt{\frac{ [B_i , B_i]}{[B_j , B_j]}}
\; , \qquad \label{corshapeamp} \\
\mathcal{T}(B_i,B_j) & \equiv & 1 - \sqrt{ \frac{[B_j \!-\! B_i,B_j \!-\! 
B_i]}{ [B_j , B_j]}} \; . \label{cortotal}
\end{eqnarray}
We evaluated all three correlators to compare our approximation (as $B_i$) 
with the simpler approximation (as $B_j$) of Adshead, Dvorkin, Hu and Peiris 
\cite{Adshead:2011bw} over the narrow range $170.8 < n_i < 173$ of the 
greatest response. The results are,
\begin{equation}
\mathcal{S} \simeq 0.9578 \qquad , \qquad \mathcal{A} \simeq 1.3436 
\qquad , \qquad \mathcal{T} \simeq 0.5189 \; . \label{usvsthem}
\end{equation}
Even though the equilateral triangle case shown by Figure~\ref{Compare} 
seems to roughly agree we can see there is quite a large mismatch in the
amplitudes that leads to a substantial degradation of the total correlator.

\section{The Square Well Model}

In 1992 Starobinsky proposed a simple model in which the first slow roll
parameter makes an instantaneous jump from one value to another, which
permits the mode functions to be solved exactly \cite{Starobinsky:1992ts}. 
Because the fundamental source of non-Gaussianity $\partial_n \ln[\epsilon(n)]$ 
is a delta function for this case, one can exactly compute $B_{4+7}(k_1,k_2,k_3)$
and derive excellent approximations for the remaining contributions
\cite{Arroja:2011yu,Martin:2011sn,Arroja:2012ae}. We shall make a slight
modification of this model in which $\epsilon(n)$ returns to its original
value after a short number of e-foldings $\Delta n$,
\begin{equation}
\epsilon(n) = \epsilon_1 \theta(n_0 \!-\! n) + \epsilon_2 \theta(n \!-\! n_0)
\theta(n_0 \!+\! \Delta n \!-\! n) + \epsilon_1 \theta(n \!-\! n_0 \!-\!
\Delta n) \; . \label{Squaremodel}
\end{equation}
We first solve exactly for the mode functions. Next a determination is made
of the parameter values for $n_0$, $\Delta n$, $\epsilon_1$ and $\epsilon_2$ 
to cause the scalar power spectrum of this model to agree with a numerical 
determination of the Step Model power spectrum of section 3.2 over the crucial 
range $170.8 < n_k < 172.8$. After that $B_{4+7}(k_1,k_2,k_3)$ is computed
exactly, and then in the approximation of setting all small factors of 
$\epsilon$ to zero. We close by using the correlators 
(\ref{corshapeamp}-\ref{cortotal}) of Hung, Fergusson and Shellard 
\cite{Hung:2019ygc} to compare this exactly solvable model with our 
approximation, and with the simpler approximation of Adshead, Dvorkin, 
Hu and Peiris \cite{Adshead:2011bw}.

For $\epsilon(n) = \epsilon_i$ for all time then the exact mode function is,
\begin{equation}
v_i(n,k) = \sqrt{\frac{\pi}{4 \epsilon_i (1 \!-\! \epsilon_i) H a^3}} \,
H^{(1)}_{\nu_i}\Biggl( \frac{k}{(1 \!-\! \epsilon_i) H a} \Biggr) \quad , 
\quad \nu_i = \frac12 \Bigl( \frac{3 \!-\! \epsilon_i}{1 \!-\! \epsilon_i}
\Bigr) \; . \label{vconst}
\end{equation}
For the actual parameter (\ref{Squaremodel}) the mode function takes the form,
\begin{eqnarray}
\lefteqn{v(n,k) = v_1(n,k) \theta(n_0 \!-\! n) } \nonumber \\
& & \hspace{1cm} + v_B(n,k) \theta(n \!-\! n_0) \theta(n_0 \!+\! \Delta n \!-\! n) 
+ v_C(n,k)\theta(n \!-\! n_0 \!-\! \Delta n) \; , \qquad
\end{eqnarray}
where $v_B(n,k)$ and $v_C(n,k)$ are,
\begin{eqnarray}
v_B(n,k) & = & \alpha v_2(n,k) + \beta v_2^*(n,k) \; , \label{vB} \\
v_C(n,k) & = & \alpha \Bigl[\gamma v_1(n,k) \!+\! \delta v_1^*(n,k)\Bigr] + 
\beta \Bigl[\gamma v_1(n,k) \!+\! \delta v_1^*(n,k)\Bigr]^* \; . \qquad 
\label{vC}
\end{eqnarray}
The appropriate matching conditions at $n = n_0$ and $n = n_0 + \Delta n$
are the continuity of $v(n,k)$ and of the product $\epsilon(n) \times
v'(n,k)$. The coefficients $\alpha$ and $\beta$ involve the mode functions
(\ref{vconst}) and their derivatives evaluated at $n = n_0$,
\begin{equation}
\alpha = -i H a^3 \Bigl[ \epsilon_2 v_1 {v_2^*}' \!-\! \epsilon_1 v_1' v_2^*
\Bigr] \quad , \quad \beta = i H a^3 \Bigl[ \epsilon_2 v_1 v_2' \!-\! 
\epsilon_1 v_1' v_2\Bigr] \; . \label{alphabeta}
\end{equation}
The coefficients $\gamma$ and $\delta$ involve the mode functions (\ref{vconst})
and their derivatives evaluated at $n = n_0 + \Delta n$,
\begin{equation}
\gamma = -i H a^3 \Bigl[ \epsilon_1 v_2 {v_1^*}' \!-\! \epsilon_2 v_2' v_1^*
\Bigr] \quad , \quad \delta = i H a^3 \Bigl[ \epsilon_1 v_2 v_1' \!-\! 
\epsilon_2 v_2' v_1\Bigr] \; .
\label{gammadelta}
\end{equation}

\begin{figure}[H]
\includegraphics[width=6.0cm,height=4.0cm]{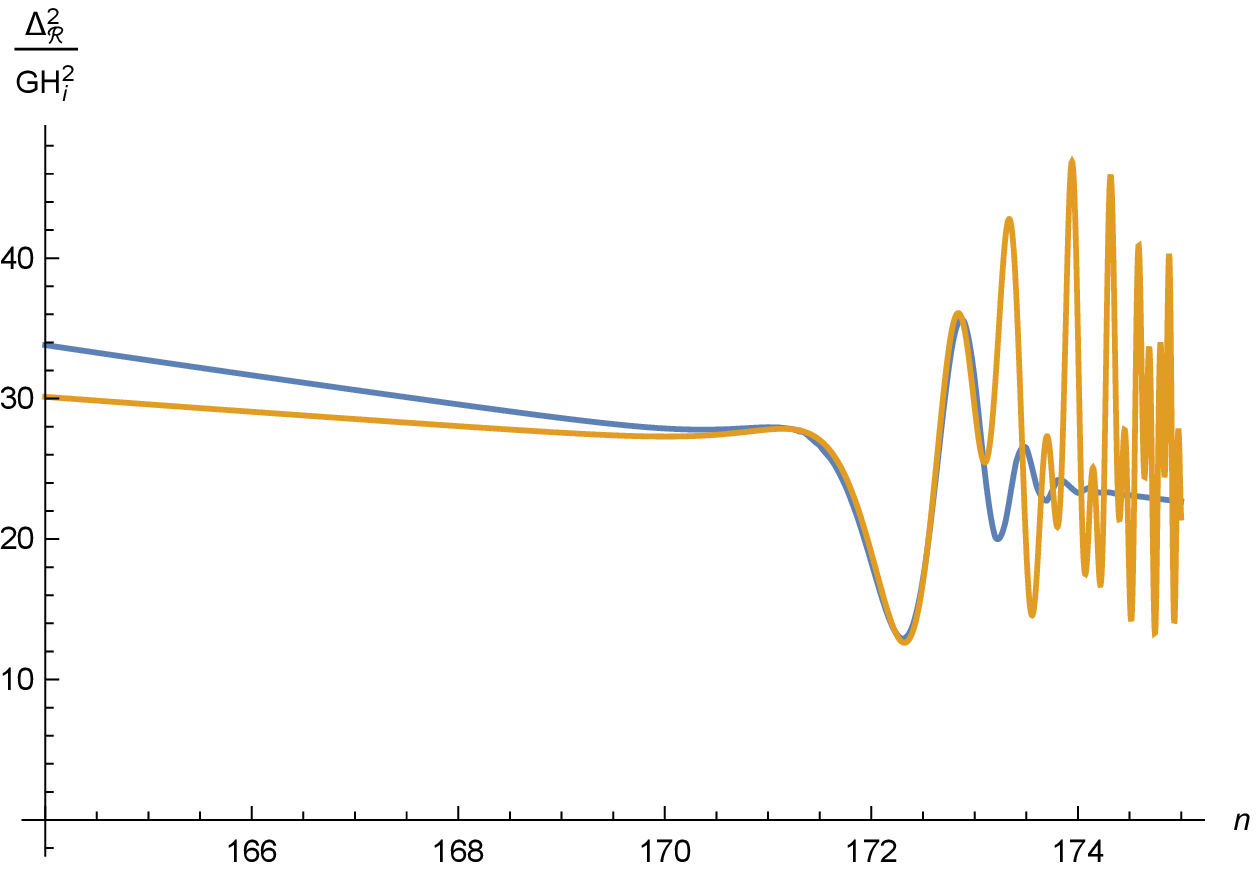}
\hspace{1cm}
\includegraphics[width=6.0cm,height=4.0cm]{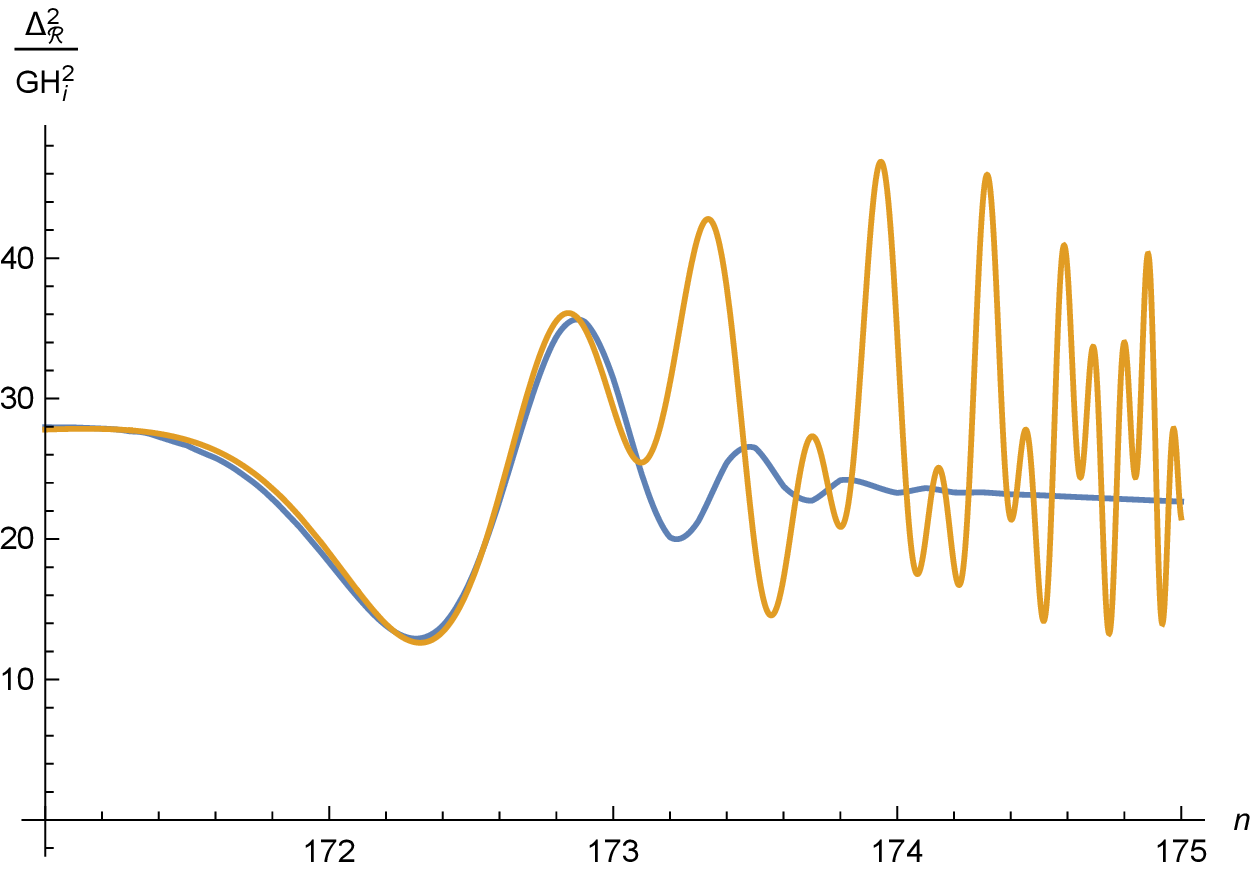}
\caption{\footnotesize Both graphs show the $\Delta^2_{\mathcal{R}}(k)$ as a 
function of e-folding of horizon crossing, for the Step Model (in blue) and for 
the best fit Square Well Model (in yellow). The approximate conversion to wave
number is $k \simeq a_i m e^{n} \sqrt{150 - \frac23 n}$, where $a_i$ is the scale
factor at the beginning of inflation and $m = 7.126 \times 10^{-6} \sqrt{8\pi G}$.}
\label{bestfit}
\end{figure}

From expression (\ref{vC}) and the small argument form of the Hankel function 
we infer the late time limit of the mode function,
\begin{equation}
\lim_{n \gg n_k} v_C(n,k) = -\frac{i H(n_k)}{\sqrt{2 \epsilon_1 k^3}}
\!\times\! \frac{\Gamma(\nu_1) [2 (1\!-\!\epsilon_1)]^{\frac1{1-\epsilon_1}}}{
\sqrt{\pi}} \!\times\! \Bigl[ \alpha (\gamma \!-\! \delta) \!-\! \beta (\gamma^* 
\!-\! \delta^*)\Bigr] \; . \label{latemode}
\end{equation}
Substituting this in expression (\ref{spectrum}) gives the Square Well model's 
prediction for the scalar power spectrum,
\begin{equation}
\Delta^2_{\mathcal{R}}(k) = \frac{G H^2(n_k)}{\pi \epsilon_1} \!\times\! 
\frac{\Gamma^2(\nu_1) [2 (1 \!-\! \epsilon_1)]^{\frac2{1-\epsilon_1}}}{\pi}
\!\times\! \Bigl\vert \alpha (\gamma \!-\! \delta) \!-\! \beta (\gamma^* \!-\!
\delta^*) \Bigr\vert^2 \; . \label{squarepower}
\end{equation}
Figure~\ref{bestfit} compares (\ref{squarepower}) with a numerical determination 
of $\Delta^2_{\mathcal{R}}(k)$ for the Step Model. There is no way to make the
two results agree for all values of $n_k$, however, very good concurrence over 
the key range of $170.8 < n_k < 172.8$ results from the following choices for 
the Square Well parameters,
\begin{equation}
n_0 = 171.3 \;\; , \; \Delta n = 0.7 \;\; , \; \epsilon_1 = 0.0093 \;\; , \;
\epsilon_2 = 0.0137 \; . \label{Squareparams}
\end{equation}
The infinite sequence of oscillations (``ringing'') evident in Fig.~\ref{bestfit}
is the result of the sharp transitions in $\epsilon(n)$ for the Square Well Model 
(\ref{Squaremodel}). For smooth transitions, such as those of the Step Model, the 
oscillations decay rapidly. Of course no one understands what caused features (if 
they are present) so it may be that the transition really {\it is} instantaneous, 
in which case ringing is a prominent signature that persists long after the 
transition. This possibility was pursued in a fascinating study by Adshead, Dvorkin, 
Hu and Lim \cite{Adshead:2011jq}. However, we shall here take the view that ringing
is an artifact of modelling smooth transitions as instantaneous, and we shall 
accordingly focus narrowly on the two e-foldings $170.8 < n_k < 172.8$ over which
the Square Well model is in reasonable agreement with the Step Model.

The great advantage of the Square Well Model is that the key modulation factor of
$\epsilon'/\epsilon$ in expression (\ref{B47}) is a delta function,
\begin{equation}
\frac{\epsilon'(n)}{\epsilon(n)} = \ln\Bigl( \frac{\epsilon_2}{\epsilon_1}\Bigr)
\Bigl[ \delta(n \!-\! n_0) - \delta(n \!-\! n_1)\Bigr] \; ,
\label{modfactor}
\end{equation}
where $n_1 \equiv n_0 + \Delta n$. We must also understand how to evaluate 
certain discontinuous factors at the jumps,
\begin{eqnarray}
\epsilon(n_0) & \!\!\!\! \longrightarrow \!\!\!\! & \Bigl( \frac{\epsilon_1 \!+\! 
\epsilon_2}{2}\Bigr) \longleftarrow \epsilon(n_1) \; , \label{lim12} \\
\epsilon(n_0) v'(n_0,k_{\alpha}) v'(n_0,k_{\beta}) & \!\!\!\! \longrightarrow 
\!\!\!\! & \Bigl( \frac{\epsilon_1 \!+\! \epsilon_2}{2}\Bigr) 
\frac{\epsilon_1}{\epsilon_2} v'_1(n_0,k_{\alpha}) v'_1(n_0,k_{\beta}) \; ,
\label{lim3} \\
\epsilon(n_1) v'(n_1,k_{\alpha}) v'(n_1,k_{\beta}) & \!\!\!\! \longrightarrow 
\!\!\!\! & \Bigl( \frac{\epsilon_1 \!+\! \epsilon_2}{2}\Bigr) 
\frac{\epsilon_2}{\epsilon_1} v'_B(n_1,k_{\alpha}) v'_B(n_1,k_{\beta}) \; . 
\label{lim4}
\end{eqnarray}
Substituting relations (\ref{modfactor}) and (\ref{lim12}-\ref{lim4}) into
expressions (\ref{Bform}) and (\ref{B47}) gives,
\begin{eqnarray}
\lefteqn{B_{4+7}(k_1,k_2,k_3) = (4\pi G)^2 \Bigl( \frac{\epsilon_1 \!+\! 
\epsilon_2}{2}\Bigr) \ln\Bigl( \frac{\epsilon_2}{\epsilon_1}\Bigr) 
{\rm Re}\Biggl[ i v_C(n_e,k_1) v_C(n_e,k_2) v_C(n_e,k_3) } \nonumber \\
& & \hspace{0cm} \times \Bigl[ H(n_0) a^3(n_0) F^*(n_0,k_1,k_2,k_3) -
H(n_1) a^3(n_1) G^*(n_1,k_1,k_2,k_3)\Bigr] \Biggr] , \qquad \label{Square47}
\end{eqnarray}
where the upper and lower factors are,
\begin{eqnarray}
\lefteqn{ F(n,k_1,k_2,k_3) \equiv \frac{(k_1^2 \!+\! k_2^2 \!+\! k_3^2)}{H^2(n) 
a^2(n)} v_1(n,k_1) v_1(n,k_2) v_1(n,k_3) } \nonumber \\
& & \hspace{0cm} - \frac{\epsilon_1}{\epsilon_2} v_1(n,k_1) v_1'(n,k_2) 
v_1'(n,k_3) - \frac{\epsilon_1}{\epsilon_2} v_1'(n,k_1) v_1(n,k_2) 
v_1'(n,k_3) \nonumber \\
& & \hspace{6.2cm} - \frac{\epsilon_1}{\epsilon_2} v_1'(n,k_1) v_1'(n,k_2) 
v_1(n,k_3) \; , \qquad \label{Fdef} \\
\lefteqn{ G(n,k_1,k_2,k_3) \equiv \frac{(k_1^2 \!+\! k_2^2 \!+\! k_3^2)}{H^2(n) 
a^2(n)} v_B(n,k_1) v_B(n,k_2) v_B(n,k_3) } \nonumber \\
& & \hspace{0cm} - \frac{\epsilon_2}{\epsilon_1} v_B(n,k_1) v_B'(n,k_2) 
v_B'(n,k_3) - \frac{\epsilon_2}{\epsilon_1} v_B'(n,k_1) v_B(n,k_2) 
v_B'(n,k_3) \nonumber \\
& & \hspace{6.2cm} - \frac{\epsilon_2}{\epsilon_1} v_B'(n,k_1) v_B'(n,k_2) 
v_B(n,k_3) \; , \qquad \label{Gdef} 
\end{eqnarray}

Expressions (\ref{Square47}-\ref{Gdef}) are exact, but somewhat opaque because 
they conceal certain large factors of $1/\epsilon$, and because they are obscured
by many other negligibly small positive powers of $\epsilon$. There is no 
appreciable loss of accuracy, and a considerable simplification, by extracting
the large factors of $1/\epsilon$ and setting the other factors of $\epsilon$ to
zero. Note that this makes the Hubble parameter constant. Two ratios which involve 
the momenta are,
\begin{equation}
\kappa_i \equiv \frac{k_i}{H(n_0) a(n_0)} \longrightarrow e^{n_{k_i} - n_0} 
\; , \; \lambda_i \equiv \frac{k_i}{H(n_1) a(n_1)} \longrightarrow 
e^{n_{k_i} - n_1} = \kappa_i e^{-\Delta n} \; . \label{kappalambda}
\end{equation}
Applying these approximations to the mode functions (at $n_0$ and $n_1$) and their 
first derivatives gives,
\begin{eqnarray}
v_i(n_0,k) \longrightarrow -\frac{iH (1 \!-\! i \kappa) e^{i\kappa}}{\sqrt{2 
\epsilon_i k^3}} & , & v_i'(n_0,k) \longrightarrow \frac{iH \kappa^2 e^{i\kappa}}{
\sqrt{2 \epsilon_1 k^3}} \; , \qquad \label{v1limit} \\
v_i(n_1,k) \longrightarrow -\frac{iH (1 \!-\! i \lambda) e^{i\lambda}}{\sqrt{2 
\epsilon_i k^3}} & , & v_i'(n_1,k) \longrightarrow \frac{iH \lambda^2 e^{i\lambda}}{
\sqrt{2 \epsilon_i k^3}} \; . \qquad \label{v2limit}
\end{eqnarray}
These approximations carry the first set of combination coefficients (\ref{alphabeta})
to,
\begin{eqnarray}
\alpha_i & \longrightarrow & \frac{i}{2 \kappa_i} \Biggl[ (1 \!-\! i\kappa_i) 
\sqrt{\frac{\epsilon_2}{\epsilon_1}} - (1 \!+\! i\kappa_i) \sqrt{\frac{\epsilon_1}{
\epsilon_2}} \, \Biggr] \; , \label{alphalimit} \\
\beta_i & \longrightarrow & \frac{i}{2 \kappa_i} (1 \!-\! i\kappa_i) \Biggl[ 
\sqrt{\frac{\epsilon_2}{\epsilon_1}} - \sqrt{\frac{\epsilon_1}{\epsilon_2}} \, \Biggr] 
e^{2 i \kappa_i} \; . \label{alphabetalimit}
\end{eqnarray}
Only the difference of the second set (\ref{gammadelta}) matters, and it becomes,
\begin{equation}
\gamma_i - \delta_i \longrightarrow \frac{e^{i \lambda_i}}{\lambda_i} \Biggl[
(1 \!-\! i \lambda_i) \sin(\lambda_i) \sqrt{\frac{\epsilon_1}{\epsilon_2}} -
\Bigl[ \sin(\lambda_i) \!-\! \lambda_i \cos(\lambda_i)\Bigr] 
\sqrt{\frac{\epsilon_2}{\epsilon_1}} \, \Biggr] \; . \label{gammadeltalimit}
\end{equation}
With these approximations expression (\ref{B47}) assumes the form,
\begin{eqnarray}
\lefteqn{B_{4+7}(k_1,k_2,k_3) \longrightarrow \frac{(\pi G H^2)^2}{k_1^2 k_2^2 k_3^2}
\Bigl( \frac{\epsilon_1 \!+\! \epsilon_2}{\epsilon_1^3}\Bigr) \ln\Bigl( 
\frac{\epsilon_2}{\epsilon_1}\Bigr) {\rm Re}\Biggl[ \frac{i A_1 A_2 A_3}{\kappa_1
\kappa_2 \kappa_3} } \nonumber \\
& & \hspace{3.5cm} \times \Bigl[ \mathcal{F}^* e^{-i (\kappa_1 + \kappa_2 + \kappa_3)} 
- \Bigl( \frac{\epsilon_1}{\epsilon_2}\Bigr)^{\frac32} \mathcal{G}^* e^{-i (\lambda_1 
+ \lambda_2 + \lambda_3)} \Bigr] \Biggr] , \qquad \label{B47limit}
\end{eqnarray}
where $A_i \equiv \alpha_i (\gamma_i - \delta_i) - \beta_i (\gamma_i^* - \delta_i^*)$ 
and the approximated factors are,
\begin{eqnarray}
\lefteqn{\mathcal{F} = (\kappa_1^2 \!+\! \kappa_2^2 \!+\! \kappa_3^2) (1 \!-\! i\kappa_1)
(1 \!-\! i \kappa_2) (1 \!-\! i \kappa_3)  } \nonumber \\
& & \hspace{2cm}- \frac{\epsilon_1}{\epsilon_2} \kappa_1^2 \kappa_2^2 (1 \!-\! i \kappa_3) 
- \frac{\epsilon_1}{\epsilon_2} \kappa_1^2 \kappa_3^2 (1 \!-\! i \kappa_2)
- \frac{\epsilon_1}{\epsilon_2} \kappa_2^2 \kappa_3^2 (1 \!-\! i \kappa_1) \; , \qquad
\label{Flimit} \\
\lefteqn{\mathcal{G} = e^{\Delta n} (\kappa_1^2 \!+\! \kappa_2^2 \!+\! \kappa_3^2) 
\prod_{i=1}^3 \Bigl[ \alpha_i (1 \!-\! i \lambda_i) - \beta_i (1 \!+\! i \lambda_i) 
e^{-2 i \lambda_i} \Bigr]} \nonumber \\
& & \hspace{0cm} - e^{-\Delta n} \frac{\epsilon_2}{\epsilon_1} \sum_{i=1}^{3} \Bigl[
\prod_{j\neq i} \kappa_j^2 (\alpha_j \!-\! \beta_j e^{-2i \lambda_j}) \Bigr]  
\Bigl[\alpha_i (1 \!-\! i \lambda_i) \!-\! \beta_i (1 \!+\! i \lambda_i)
e^{-2i \lambda_i} \Bigr] \; . \qquad \label{Glimit}
\end{eqnarray}

One can see from Figure~\ref{bestfit} that the power spectra of the Square Well Model 
and the step Model agree almost perfectly over the region $170.8 < n < 173$. This 
might seem to indicate that they would produce the nearly the same non-Gaussian signal, 
at least when restricted to the same narrow range. However, the results are disappointing
when the two models are compared using the shape, amplitude and total correlators 
(\ref{corshapeamp}-\ref{cortotal}) of Hung, Fergusson and Shellard \cite{Hung:2019ygc},
\begin{equation}
\mathcal{S} \simeq 0.7976 \qquad , \qquad \mathcal{A} \simeq 1.1050 \qquad , \qquad
\mathcal{T} \simeq 0.3230 \; , \label{SWvsSM}
\end{equation} 
where $B_i$ was the Square Well Model and $B_j$ was the Step Model. The amplitudes of
the two models are in much better agreement than for the comparison (\ref{usvsthem}) 
of the Step Model with the approximation of Adshead, Dvorkin, Hu and Peiris 
\cite{Adshead:2011bw}. However, the shapes disagree, which results in an even lower
total correlator. Note that the problem in this case did not arise from inaccurately 
modeling the non-Gaussian response to a given history $\epsilon(n)$, but rather from 
the fact that different histories produce different bi-spectra, even when the power 
spectra are very similar.

We also compared the Square Well Model (as $B_i$) with the approximation of Adshead, 
Dvorkin, Hu and Peiris \cite{Adshead:2011bw} (as $B_j$),
\begin{equation}
\mathcal{S} \simeq 0.8946 \qquad , \qquad \mathcal{A} \simeq 1.4847 \qquad , \qquad
\mathcal{T} \simeq 0.2598 \; . \label{SWvsADHP}
\end{equation}
Both the shape correlator and the amplitude correlator are worse than for the 
comparison (\ref{usvsthem}) of our approximation with that of Adshead, Dvorkin, Hu 
and Peiris, resulting in a much smaller total correlator.

\section{Epilogue}

We have examined the non-Gaussianity associated with conjectured sharp variations 
in the first slow roll parameter $\epsilon(n)$ known as ``features''. In section 2 
we identified the crucial contribution, equation (\ref{B47}), which becomes significant 
for features. Section 3 applied an approximation for how the scalar mode functions 
depend analytically on $\epsilon(n)$ \cite{Brooker:2017kjd,Brooker:2017kij} to 
develop an approximation (\ref{approxB47}) for this term. Our result involves
three tabulated functions of the instantaneous e-folding $n$ and the e-folding
of horizon crossing $n_k$:
\begin{enumerate}
\item{$\widetilde{g}(n,n_k)$ given in expression (\ref{gapprox});}
\item{$\widetilde{\gamma}'(n,n_k)$ given in expression (\ref{gprime}); and}
\item{$\widetilde{\phi}(n,n_k)$ given in expression (\ref{phiapprox}).}
\end{enumerate}
Although generating these functions is numerically challenging, it only needs to be
done over the narrow range of $n$ and $n_k$ associated with the feature. This is
illustrated in Figure~\ref{ModelDeviations} which identifies the small ranges of
$n$ and $n_k$ over which significant corrections would occur for a model of the
first feature.

Our technique is more time-consuming, but also more accurate, than the
approximation of Adshead, Dvorkin, Hu and Peiris \cite{Adshead:2011bw}. When the
two approximations were compared using the total correlator (\ref{cortotal})
of Hung, Fergusson and Shellard \cite{Hung:2019ygc} the result (\ref{usvsthem})
was nearly a 50\% degradation of the signal, even when the comparison was 
restricted to a narrow range around the feature. Accurate modeling is crucial
when studying features because they produce an oscillating signal, so that even 
small errors in the phase can significantly degrade the signal. This is especially 
relevant because the response to a feature is delayed to later crossing wave 
numbers. Unless the late time phase information is accurately modeled, trying to 
boost the signal by including the delayed response will actually reduce the 
measured signal.

In section 4 we presented a slight elaboration of a model due to Starobinsky 
\cite{Starobinsky:1992ts} for which the crucial contribution (\ref{B47}) can
be computed exactly, without any approximation \cite{Arroja:2011yu,Martin:2011sn,
Arroja:2012ae}. In our model $\epsilon(n)$ jumps from $\epsilon_1$ to $\epsilon_2$
and then falls back down after an interval $\Delta n$, hence the name ``Square Well 
Model''. Expression (\ref{Square47}) gives the exact result for the bi-spectrum of
the Square Well Model. However, taking the inessential factors of $\epsilon$ to zero 
produces a simpler and more transparent result (\ref{B47limit}) which is almost as
accurate. A consequence of the sharp transitions is the persistence of oscillations
for wave numbers which experience horizon crossing long after the transition. We
regarded this as an artifact of the square well approximation, and truncated the 
late oscillations. For a different point of view we recommend the study of Adshead, 
Dvorkin, Hu and Lim \cite{Adshead:2011jq}.

Figure~\ref{Compare} shows that the power spectra of the Square Well Model agree 
with that of the Step Model over the narrow range of $170.8 < n < 173$. However,
the bi-spectra they produce are very different. We found a total shape correlator 
(\ref{SWvsSM}) of only about one third! This underlines the importance of knowing 
the history $\epsilon(n)$ in addition to accurately modeling the response to it.

\vspace{.5cm}

\centerline{\bf Acknowledgements}

This work was partially supported by the European Union's Seventh Framework 
Programme (FP7-REGPOT-2012-2013-1) under grant agreement number 316165; by 
the European Union's Horizon 2020 Programme under grant agreement 
669288-SM-GRAV-ERC-2014-ADG; by NSF grants PHY-1506513, 1806218 and 1912484; 
and by the UF's Institute for Fundamental Theory.


\begin{thebibliography}{99}

\bibitem{Mukhanov:1981xt} 
  V.~F.~Mukhanov and G.~V.~Chibisov,
  JETP Lett.\  {\bf 33}, 532 (1981)
  [Pisma Zh.\ Eksp.\ Teor.\ Fiz.\  {\bf 33}, 549 (1981)].
  
\bibitem{Woodard:2009ns} 
  R.~P.~Woodard,
  Rept.\ Prog.\ Phys.\  {\bf 72}, 126002 (2009)
  doi:10.1088/0034-4885/72/12/126002
  [arXiv:0907.4238 [gr-qc]].
    
\bibitem{Ashoorioon:2012kh} 
  A.~Ashoorioon, P.~S.~Bhupal Dev and A.~Mazumdar,
  Mod.\ Phys.\ Lett.\ A {\bf 29}, no. 30, 1450163 (2014)
  doi:10.1142/S0217732314501636
  [arXiv:1211.4678 [hep-th]].

\bibitem{Krauss:2013pha} 
  L.~M.~Krauss and F.~Wilczek,
  Phys.\ Rev.\ D {\bf 89}, no. 4, 047501 (2014)
  doi:10.1103/PhysRevD.89.047501
  [arXiv:1309.5343 [hep-th]].
  
\bibitem{Aghanim:2018eyx} 
  N.~Aghanim {\it et al.} [Planck Collaboration],
  arXiv:1807.06209 [astro-ph.CO].

\bibitem{Array:2015xqh} 
  P.~A.~R.~Ade {\it et al.} [BICEP2 and Keck Array Collaborations],
  Phys.\ Rev.\ Lett.\  {\bf 116}, 031302 (2016)
  doi:10.1103/PhysRevLett.116.031302
  [arXiv:1510.09217 [astro-ph.CO]].

\bibitem{Tsamis:1997rk} 
  N.~C.~Tsamis and R.~P.~Woodard,
  Annals Phys.\  {\bf 267}, 145 (1998)
  doi:10.1006/aphy.1998.5816
  [hep-ph/9712331].

\bibitem{Saini:1999ba} 
  T.~D.~Saini, S.~Raychaudhury, V.~Sahni and A.~A.~Starobinsky,
  Phys.\ Rev.\ Lett.\  {\bf 85}, 1162 (2000)
  doi:10.1103/PhysRevLett.85.1162
  [astro-ph/9910231].

\bibitem{Padmanabhan:2002cp} 
  T.~Padmanabhan,
  Phys.\ Rev.\ D {\bf 66}, 021301 (2002)
  doi:10.1103/PhysRevD.66.021301
  [hep-th/0204150].
  
\bibitem{Nojiri:2005pu} 
  S.~Nojiri and S.~D.~Odintsov,
  Gen.\ Rel.\ Grav.\  {\bf 38}, 1285 (2006)
  doi:10.1007/s10714-006-0301-6
  [hep-th/0506212].
  
\bibitem{Woodard:2006nt} 
  R.~P.~Woodard,
  Lect.\ Notes Phys.\  {\bf 720}, 403 (2007)
  doi:10.1007/978-3-540-71013-4\_14
  [astro-ph/0601672].
  
\bibitem{Guo:2006ab} 
  Z.~K.~Guo, N.~Ohta and Y.~Z.~Zhang,
  Mod.\ Phys.\ Lett.\ A {\bf 22}, 883 (2007)
  doi:10.1142/S0217732307022839
  [astro-ph/0603109].
  
\bibitem{Ijjas:2013vea} 
  A.~Ijjas, P.~J.~Steinhardt and A.~Loeb,
  Phys.\ Lett.\ B {\bf 723}, 261 (2013)
  doi:10.1016/j.physletb.2013.05.023
  [arXiv:1304.2785 [astro-ph.CO]].

\bibitem{Guth:2013sya} 
  A.~H.~Guth, D.~I.~Kaiser and Y.~Nomura,
  Phys.\ Lett.\ B {\bf 733}, 112 (2014)
  doi:10.1016/j.physletb.2014.03.020
  [arXiv:1312.7619 [astro-ph.CO]].
  
\bibitem{Linde:2014nna} 
  A.~Linde,
  doi:10.1093/acprof:oso/9780198728856.003.0006
  arXiv:1402.0526 [hep-th].
  
\bibitem{Ijjas:2014nta} 
  A.~Ijjas, P.~J.~Steinhardt and A.~Loeb,
  Phys.\ Lett.\ B {\bf 736}, 142 (2014)
  doi:10.1016/j.physletb.2014.07.012
  [arXiv:1402.6980 [astro-ph.CO]].
  
\bibitem{Ade:2015lrj} 
  P.~A.~R.~Ade {\it et al.} [Planck Collaboration],
  Astron.\ Astrophys.\  {\bf 594}, A20 (2016)
  doi:10.1051/0004-6361/201525898
  [arXiv:1502.02114 [astro-ph.CO]].
  
\bibitem{Torrado:2016sls} 
  J.~Torrado, B.~Hu and A.~Achucarro,
  Phys.\ Rev.\ D {\bf 96}, no. 8, 083515 (2017)
  doi:10.1103/PhysRevD.96.083515
  [arXiv:1611.10350 [astro-ph.CO]].
  
\bibitem{Martin:2006rs} 
  J.~Martin and C.~Ringeval,
  JCAP {\bf 0608}, 009 (2006)
  doi:10.1088/1475-7516/2006/08/009
  [astro-ph/0605367].
  
\bibitem{Covi:2006ci} 
  L.~Covi, J.~Hamann, A.~Melchiorri, A.~Slosar and I.~Sorbera,
  Phys.\ Rev.\ D {\bf 74}, 083509 (2006)
  doi:10.1103/PhysRevD.74.083509
  [astro-ph/0606452].
  
\bibitem{Hamann:2007pa} 
  J.~Hamann, L.~Covi, A.~Melchiorri and A.~Slosar,
  Phys.\ Rev.\ D {\bf 76}, 023503 (2007)
  doi:10.1103/PhysRevD.76.023503
  [astro-ph/0701380].
  
\bibitem{Hazra:2014goa}
  D.~K.~Hazra, A.~Shafieloo, G.~F.~Smoot and A.~A.~Starobinsky,
  JCAP {\bf 1408} (2014) 048
  doi:10.1088/1475-7516/2014/08/048
  [arXiv:1405.2012 [astro-ph.CO]].

\bibitem{Hazra:2016fkm} 
  D.~K.~Hazra, A.~Shafieloo, G.~F.~Smoot and A.~A.~Starobinsky,
  JCAP {\bf 1609}, no. 09, 009 (2016)
  doi:10.1088/1475-7516/2016/09/009
  [arXiv:1605.02106 [astro-ph.CO]].
  
\bibitem{Achucarro:2012fd} 
  A.~Achúcarro, J.~O.~Gong, G.~A.~Palma and S.~P.~Patil,
  Phys.\ Rev.\ D {\bf 87}, no. 12, 121301 (2013)
  doi:10.1103/PhysRevD.87.121301
  [arXiv:1211.5619 [astro-ph.CO]].

\bibitem{Brooker:2016imi} 
  D.~J.~Brooker, N.~C.~Tsamis and R.~P.~Woodard,
  Phys.\ Lett.\ B {\bf 773}, 225 (2017)
  doi:10.1016/j.physletb.2017.08.027
  [arXiv:1603.06399 [astro-ph.CO]].
  
\bibitem{Brooker:2017kjd} 
  D.~J.~Brooker, N.~C.~Tsamis and R.~P.~Woodard,
  Phys.\ Rev.\ D {\bf 96}, no. 10, 103531 (2017)
  doi:10.1103/PhysRevD.96.103531
  [arXiv:1708.03253 [gr-qc]].
                                
\bibitem{Maldacena:2002vr} 
  J.~M.~Maldacena,
  JHEP {\bf 0305}, 013 (2003)
  doi:10.1088/1126-6708/2003/05/013
  [astro-ph/0210603].
  
\bibitem{Fergusson:2006pr} 
  J.~R.~Fergusson and E.~P.~S.~Shellard,
  Phys.\ Rev.\ D {\bf 76}, 083523 (2007)
  doi:10.1103/PhysRevD.76.083523
  [astro-ph/0612713].
  
\bibitem{Fergusson:2008ra} 
  J.~R.~Fergusson and E.~P.~S.~Shellard,
  Phys.\ Rev.\ D {\bf 80}, 043510 (2009)
  doi:10.1103/PhysRevD.80.043510
  [arXiv:0812.3413 [astro-ph]].
  
\bibitem{Ade:2015ava} 
  P.~A.~R.~Ade {\it et al.} [Planck Collaboration],
  Astron.\ Astrophys.\  {\bf 594}, A17 (2016)
  doi:10.1051/0004-6361/201525836
  [arXiv:1502.01592 [astro-ph.CO]].

\bibitem{Adshead:2011bw} 
  P.~Adshead, W.~Hu, C.~Dvorkin and H.~V.~Peiris,
  Phys.\ Rev.\ D {\bf 84}, 043519 (2011)
  doi:10.1103/PhysRevD.84.043519
  [arXiv:1102.3435 [astro-ph.CO]].

\bibitem{Adshead:2011jq} 
  P.~Adshead, C.~Dvorkin, W.~Hu and E.~A.~Lim,
  Phys.\ Rev.\ D {\bf 85}, 023531 (2012)
  doi:10.1103/PhysRevD.85.023531
  [arXiv:1110.3050 [astro-ph.CO]].
    
\bibitem{Hazra:2012yn} 
  D.~K.~Hazra, L.~Sriramkumar and J.~Martin,
  JCAP {\bf 1305}, 026 (2013)
  doi:10.1088/1475-7516/2013/05/026
  [arXiv:1201.0926 [astro-ph.CO]].
  
\bibitem{Brooker:2017kij} 
  D.~J.~Brooker, N.~C.~Tsamis and R.~P.~Woodard,
  JCAP {\bf 1804}, no. 04, 003 (2018)
  doi:10.1088/1475-7516/2018/04/003
  [arXiv:1712.03462 [gr-qc]].
  
\bibitem{Liddle:1994dx} 
  A.~R.~Liddle, P.~Parsons and J.~D.~Barrow,
  Phys.\ Rev.\ D {\bf 50}, 7222 (1994)
  doi:10.1103/PhysRevD.50.7222
  [astro-ph/9408015].
  
\bibitem{Mukhanov:1990me} 
  V.~F.~Mukhanov, H.~A.~Feldman and R.~H.~Brandenberger,
  Phys.\ Rept.\  {\bf 215}, 203 (1992).
  doi:10.1016/0370-1573(92)90044-Z
  
\bibitem{Liddle:1993fq} 
  A.~R.~Liddle and D.~H.~Lyth,
  Phys.\ Rept.\  {\bf 231}, 1 (1993)
  doi:10.1016/0370-1573(93)90114-S
  [astro-ph/9303019].
  
\bibitem{Salopek:1988qh} 
  D.~S.~Salopek, J.~R.~Bond and J.~M.~Bardeen,
  Phys.\ Rev.\ D {\bf 40}, 1753 (1989).
  doi:10.1103/PhysRevD.40.1753

\bibitem{Romania:2012tb} 
  M.~G.~Romania, N.~C.~Tsamis and R.~P.~Woodard,
  JCAP {\bf 1208}, 029 (2012)
  doi:10.1088/1475-7516/2012/08/029
  [arXiv:1207.3227 [astro-ph.CO]].
  
\bibitem{Adams:2001vc} 
  J.~A.~Adams, B.~Cresswell and R.~Easther,
  Phys.\ Rev.\ D {\bf 64}, 123514 (2001)
  doi:10.1103/PhysRevD.64.123514
  [astro-ph/0102236].

\bibitem{Mortonson:2009qv} 
  M.~J.~Mortonson, C.~Dvorkin, H.~V.~Peiris and W.~Hu,
  Phys.\ Rev.\ D {\bf 79}, 103519 (2009)
  doi:10.1103/PhysRevD.79.103519
  [arXiv:0903.4920 [astro-ph.CO]].

\bibitem{Hung:2019ygc} 
  J.~Hung, J.~R.~Fergusson and E.~P.~S.~Shellard,
  arXiv:1902.01830 [astro-ph.CO].
  
\bibitem{Starobinsky:1992ts} 
  A.~A.~Starobinsky,
  JETP Lett.\  {\bf 55}, 489 (1992)
  [Pisma Zh.\ Eksp.\ Teor.\ Fiz.\  {\bf 55}, 477 (1992)].

\bibitem{Arroja:2011yu} 
  F.~Arroja, A.~E.~Romano and M.~Sasaki,
  Phys.\ Rev.\ D {\bf 84}, 123503 (2011)
  doi:10.1103/PhysRevD.84.123503
  [arXiv:1106.5384 [astro-ph.CO]].
  
\bibitem{Martin:2011sn} 
  J.~Martin and L.~Sriramkumar,
  JCAP {\bf 1201}, 008 (2012)
  doi:10.1088/1475-7516/2012/01/008
  [arXiv:1109.5838 [astro-ph.CO]].

\bibitem{Arroja:2012ae} 
  F.~Arroja and M.~Sasaki,
  JCAP {\bf 1208}, 012 (2012)
  doi:10.1088/1475-7516/2012/08/012
  [arXiv:1204.6489 [astro-ph.CO]].
      
\end{thebibliography}
\end{document}